\documentclass[journal=acsodf,manuscript=article]{achemso}
\usepackage{stmaryrd}
\usepackage{amsmath}
\usepackage{amssymb}
\usepackage{graphics}
\usepackage{subfigure}
\usepackage{fullpage}
\usepackage{mciteplus}
\usepackage{color}
\usepackage{hyperref}
\usepackage{makecell}
\newcommand{\ka}{k_{\mbox{\scriptsize on}}}
\newcommand{\km}{k_{\mbox{\scriptsize off}}}
\newcommand{\alc}[1]{\alpha_{c#1}}
\newcommand{\alh}[1]{\alpha_{h#1}}
\newcommand{\alco}[1]{\alpha^{(obs)}_{c#1}} 
\newcommand{\alho}[1]{\alpha^{(obs)}_{h#1}}

\newcommand{\dalc}[1]{\alpha^{(1)}_{c#1}}
\newcommand{\dalh}[1]{\alpha^{(1)}_{h#1}}
\newcommand{\dal}[1]{\alpha^{(1)}_{#1}}
\newcommand{\halc}[1]{\hat{\alpha}_{c#1}}
\newcommand{\halh}[1]{\hat{\alpha}_{h#1}}
\newcommand{\hal}[1]{\hat{\alpha}_{#1}}
\newcommand{\hdalc}[1]{\hat{\alpha}^{(1)}_{c#1}}
\newcommand{\hdalh}[1]{\hat{\alpha}^{(1)}_{h#1}}
\newcommand{\hdal}[1]{\hat{\alpha}^{(1)}_{#1}}
\newcommand{\kr}[1]{\frac{1}{h}K\left(\frac{#1}{h}\right)}
\title{A Unified Statistical Procedure to Analyse Irreversible Thermal Curves}

\author{Sanjay Chaudhuri}
\affiliation{Department of Statistics, University of Nebraska-Lincoln, Lincoln, NE 68583, USA}

\author{Dang Trung Kien}
\affiliation{Department of Biostatistics,
  Saw Swee Hock School of Public Health,
  National University of Singapore}

\author{Gopinatha Suresh Kumar}
\affiliation{Biophysical Chemistry Laboratory, CSIR-Indian Institute of Chemical Biology, 4 Raja S. C. Mullick Road, Jadavpur, Kolkata 700 032, India}
\author{Souvik Maiti}
\affiliation{Proteomics and Structural Biology Unit, Institute of Genomics and Integrative Biology, CSIR, Mall Road, Delhi, 110 007, India}
\author{Daisuke Miyoshi}
\affiliation{Frontiers of Innovative Research in Science and Technology (FIRST), Konan University, 7-1-20 Minatojima-minamimachi, Chuo-Ku,Kobe 650-0047, Japan}

\author{Jhimli Bhattacharyya}
\email{jhimli@nitnagaland.ac.in}
\affiliation{Department of Chemistry, National Institute of Technology Nagaland, Chumukedima, Dimapur, Nagaland, 797 103, India.}

\begin{document}
\begin{abstract}
  DNA hybridisation experiments are crucial for studying the thermodynamic and kinetic profiles of various systems in nucleic acid chemistry.  The phenomenon of hysteresis is commonly observed in many such UV thermal experiments involving unmodified or modified nucleic acids.  In the presence of hysteresis, the thermal curves are irreversible and demand a significant effort to produce the reaction-specific kinetic and thermodynamic parameters.  In this article, we describe a unified statistical procedure to analyse such thermal curves.
  More specifically, the proposed method allows one to handle the thermal curves for the formation of duplexes, triplexes, and various quadruplexes in exactly the same way.  
The proposed method uses a local polynomial regression to find the smoothed thermal curves and calculate their slopes.  This method is more flexible and easier to implement than the least squares polynomial smoothing, which is currently almost universally used for such purposes.  
Full analyses of the curves, including computation of kinetic and thermodynamic parameters, can be done using freely available statistical software.  The proposed procedure has been implemented in a web-based free software called ``anhysnuc'', which can be found at \href{https://sanjaychaudhuri.shinyapps.io/anhysnuc/}{https://sanjaychaudhuri.shinyapps.io/anhysnuc/}.
Finally, we illustrate our method by analysing irreversible curves encountered in the formation of a G-quadruplex and an LNA-modified parallel duplex.

\end{abstract}

\maketitle

\section{Introduction}
Analysis of experimental data is an integral part of chemical research.  Even if a chemical experiment is well-designed and extremely well-executed, no conclusions can be drawn from it unless the results are properly analyzed. In fact, erroneous analysis of experimental data can lead to wrong conclusions, resulting in a huge waste of effort and resources. 

Data obtained from chemical experiments require several advanced statistical tools for a proper analysis.  Each of these steps is computationally intensive and demands advanced mathematical and statistical expertise from a researcher.  Many procedures require tedious calculations, too difficult and time-consuming to perform without advanced computational facilities.  Many researchers use computational tools designed specifically for their own needs.\citep{jfit,miyoshi:et:al:2006,miyoshi:et:al:2009,ray:maiti:nandy:1996,mirzoev:lyubartsev:2013} 
However, in most cases, no statistical software is freely available.  Many commercial softwares focus only on a part of the task and different softwares are required to complete the procedure.  Lack of analytic tools often restricts choices at the design stage of the experiment and more often inflates the cost and effort even with a loss of accuracy.     



Thermodynamic and kinetic analyses of hybridization reactions involving various unmodified and modified nucleic acids have been a topic of immense interest among researchers in biophysical chemistry for at least half a century.  Such investigations can be helpful in determining the stability of the secondary structures in nucleic acids.  They can also be used to study the hybridization between an oligonucleotide and a specific target.
Even though the thermodynamics of such experiments are well established, their kinetic properties are complex and often less understood \citep{ashwood:tomakoff:2023,mukherjee:purakayastha:2024,halder:purakayasha:2022}.  
Results from these studies are useful for PCR primer design, gene sequencing and a host of bio-medical applications \citep{mehra:kaushik:2023,sharma:pandya:2023,karan:hossain:2025,zhuLiuEtal2025,kumarGhoshEtal2024,rayGhosh2025,mondalGao2021}.

Several procedures for determining thermodynamic and kinetic parameters of a hybridisation experiment are available. One can use techniques like differential scanning calorimetry (DSC), isothermal titration calorimetry (ITC), fluorescence spectroscopy, circular dichroism (CD), etc \citep{dhal:hossain:2023,bag:bhowmik:2025,mandal:pramanik:2025}.  Another popular and convenient way to determine such parameters is through the analysis of thermal curves obtained by UV-Vis spectrophotometry. In general, for the biophysical characterisation of nucleic acid reactions, UV thermal experiments are quite effective. 
For nucleic acids, denaturation (heating/melting) and renaturation (cooling/annealing) processes are usually reversible. \citep{mergny:lacroix:2003} That is, if a sample of denatured nucleic acid is cooled, in most cases renaturation of the structure is observed.
Thermodynamic parameters can be obtained by analyzing such thermal curves, \citep{jfit,miyoshi:et:al:2006,miyoshi:et:al:2009,mergny:lacroix:2003,sugimoto:et:al:2001,rougee:et:al:1992} provided the denaturation (melting) and renaturation (annealing) curves can be superimposed on each other (\emph{i.e} the folding and unfolding rate constants of the nucleic acids are equal).
With many modified and unmodified oligonucleotides, however, such is not the case.  For various reasons, the renaturation process may be quite slow.\citep{mergny:lacroix:2003,cantor:schimmel,brown:rachwal:fox:2005}  
As a result, if the temperature gradient during the melting and annealing is not slow enough,  the heating and cooling curves are not superimposed and thus none of them correspond to the true equilibrium curve. \citep{mergny:lacroix:2003} This phenomenon is known as hysteresis.

\textcolor{black}{In nucleic acid hybridization, hysteresis primarily occurs when annealing happens at a much slower rate than melting.\citep{cantor:schimmel}  In practice, nucleic acids don’t just melt or anneal instantaneously. Often, depending on the speed of heating-cooling, and the nature of the base sequences (loop or strand size/length, composition), etc., the system can get trapped in intermediate metastable states.  Getting out of such states requires overcoming energy barriers, which slow down, especially the annealing process. This produces non-overlapping melting and annealing curves. In some cases, the speed of heating and cooling can significantly impact hysteresis.  It is believed that a fast rate of change in temperature does not allow enough time for the system to reach perfect equilibrium. Furthermore, there is a hypothetical condition under which the equilibrium can be reached, where hysteresis will disappear.  Determining such experimental conditions is usually tedious, non-implementable in practice, and may waste considerable time and resources.}

Hysteresis has been discussed by several authors before. \citet{anshelevich:1984} perform a theoretical study of melting processes where hysteresis is observed.  A classic example can be seen during the fast melting and annealing of calf thymus DNA. \citep{cantor:schimmel} 
In recent times, such a phenomenon has been reported from the study of RNA and DNA binding to a pyrrolidine-amide oligonucleotide mimic (POM) \citep{hickman:micklefield:2000}.

\textcolor{black}{An important example with vast bio-medical applications can be found in the \emph{vide infra} hybridization experiments involving locked nucleic acid (LNA) modified duplexes.\citep{bhattacharyya:et:al:2011}  Locked nucleic acids or bridged nucleic acids are a well-known group of chemically modified nucleotides. \citep{singh:et:al:1998,obika:et:al:1998} LNA-protected DNA analogs exhibit enhanced thermal stability when hybridized with complementary strands and their thermal stability often increases on successive LNA modifications. \citep{koshkin:et:al:1998,singh:wengel:1998,wengel:1998,wengel:et:al:2001,Wahlestedt:et:al:2000, bhattacharyya:et:al:2011,maiti:et:al:2006,maiti:et:al:2008}}

\textcolor{black}{Many authors such as} \citep{kaur:babu:maiti:2007,bhattacharyya:et:al:2011} \citet{rougee:et:al:1992} reported and analyzed non-equilibrium thermal curves in triplex formation.  
It has also been observed in the triplex formation of DNA oligomers with Guanidinium and methylthiourea-link nucleosides. \citep{blasko:bruice:1996,arya:bruice:1999}  Same phenomenon was seen in triplex formation with $\alpha$-L-LNA \citep{maiti:2006} and peptide nucleic acids (PNA). \citep{lesnik:freier:1997}  It is known that the melting of tetramolecular DNA and RNA quadruplexes is not kinetically reversible. \citep{mergny:et:al:2005,tran:mergny:2012} However, for others, specially the intramolecular G-quadruplexes hysteresis is commonly observed. \citep{mergny:lacroix:2009,mergny:lacroix:2003,brown:rachwal:fox:2005,sacca:mergny:2005,rachwalEtAl2007,rachwalFox2007,rachwalBrownFox2007,zhang:balasubramanian:2012,pandey:maiti:2013,sekiboFox2017} 


If hysteresis is observed, the thermodynamic parameters cannot be estimated directly from the thermal curves.  Nevertheless, it is possible to observe a simple one-way reaction from a folded state to an unfolded state with a melting curve, whereas the annealing curve traces the transition from an unfolded random coil to a folded hybrid.  
Thus, if hysteresis is observed, kinetic analysis of melting and annealing curves can be attempted. \citep{bhattacharyya:et:al:2011} The annealing and melting curves respectively allow one to determine the association and dissociation rate constants.  Assuming a two-state (all or none) model, these kinetic rate constants yield the corresponding activation energies. From these rate constants and the activation energies, thermodynamic parameters, \emph{e.g.} changes in free energy, enthalpy or entropy can be obtained.  The analysis mentioned above was first described by \citet{rougee:et:al:1992} and has been successfully used for kinetic analysis of many non-canonical structures. However, as we shall demonstrate below, it is pretty laborious to perform this kinetic analysis manually.  
First of all, the raw experimental data has to be smoothed and the value of the derivative has to be accurately calculated at each observation. The values of rate constants at each temperature are obtained by solving a pair of simultaneous equations.  The activation energies are calculated from the rates of changes in the two rate constants.  To our knowledge, no software which tackles the whole procedure is available. 
In most cases if the UV thermal curves show hysteresis, estimation of the kinetic parameters, activation energies, and hence the thermodynamic profile become extremely time-consuming.  Thus in such cases, thermodynamic analysis is often not attempted from the spectrophotometric data.  
However, a thorough thermodynamic profile is necessary to understand the chemical interactions of nucleic acids. This thermodynamic profile complements the structural data providing a clear and coherent picture.  \citep{kaur:babu:maiti:2007,maiti:et:al:2008,sugimoto:et:al:2001,gsk:2011,bhadra:gsk:2011,saha:hossain:gsk:2010}


In this article, we describe a unified statistical procedure for kinetic analysis of UV thermal curves showing hysteresis.  Our procedure is available as a freely available web-based software called ``anhysnuc''.  The application uses freely available libraries and codes from a statistical package called {\sf R} \citep{R_manual}.  Anhysnuc can be found at \href{https://sanjaychaudhuri.shinyapps.io/anhysnuc/}{https://sanjaychaudhuri.shinyapps.io/anhysnuc/}.
First, local polynomial regression methods are used to smooth the raw experimental data and estimate the derivative of the underlying thermal curve at each temperature.  The application then calculates the kinetic rate constants,
  activation energies, and thermodynamic parameters using various statistical methods. The numerical results, including the figures, can be easily downloaded for subsequent use.

We illustrate our procedure with two data from four sets of thermal curves.  The first two sets consider intramolecular G-quadruplex formation of two specific RNA strands.  Stability of these nucleic acids was investigated by \citet{pandey:maiti:2013}, who however, in order to avoid hysteresis performed the experiment with a flatter temperature gradient than the data used here. Next, we use data reported by \citet{bhattacharyya:et:al:2011} where kinetic and thermodynamic effects of
LNA modification on parallel and antiparallel DNA duplexes was reported.  Significant hysteresis was observed for LNA-modified parallel duplexes for all temperature gradients used.  The extent of hysteresis increased with the increase in the number of LNA modifications.  Our results closely match with those reported in the original references.


\section{UV Spectrophotometric Experiment to Obtain the Thermal Curves}

The UV absorbance spectra are recorded on a spectrophotometer equipped with a temperature controller. The heating and cooling rates are usually kept constant throughout the experiment.  The sample is usually first annealed from a high ($95^{\circ}$ C) to a low ($0^{\circ}$ C) temperature at a steady rate.  The annealed sample is then incubated at a low temperature for some time. When the temperature is below $20^{\circ}$C the cell chamber is flushed with dry nitrogen gas to avoid condensation. Then at the melting stage, the temperature is steadily increased from a low to a high value.  The annealing and the melting curves are usually obtained by measuring the UV absorbance at $260$ or $295$ nm. 
A buffer is used to control the pH of the system. The final data is obtained by subtracting the absorbance due to the buffer from the observed data.   

\section{The Theory Behind the Kinetic and Thermodynamic Analysis of Thermal Curves Exhibiting Hysteresis}\label{sec:theory}

In presence of hysteresis, for a kinetic analysis of the thermal curves one uses a two-state model between the reactant ($R$) and the product ($P$).  Many situations are possible in this case. For example, a single strand can fold into a duplex or a more complex hybrid.  
The reaction may be bimolecular involving either two autocomplimentary oligonucleotides or two different complimentary oligonucleotides. The kinetic order of the reaction differs from one group to another.  
We consider two broad classes of reactions discussed in Table $2$ of \citet{mergny:lacroix:2003}, \textcolor{black}{namely, the intramolecular reactions, and the intermolecular reactions with two complimentary oligonucleotides}, below.  Pronounced hysteresis have been observed by several authors, for reactions belonging to these classes. 

\subsubsection{Intramolecular reactions}  For intramolecular reactions, a monomeric nucleic acid strand ($R$) folds to produce various hybrid structures  ($P$). Hairpin duplex, intramolecular triplex, intramolecular G-quadruplex formations are examples belonging to this class. 

At equilibrium, the reaction can be written as:
\begin{equation}\label{eq:intra}
R\leftrightarrows P.
\end{equation}
Suppose $\ka$ denotes the rate constant of annealing or cooling, which means association and $\km$ is the rate constant of melting or heating which results in dissociation.  From \eqref{eq:intra} the rate of change of concentration of the product $P$ at time $t$ is given by
\begin{equation}\label{eq:intrap}
\frac{d[P]}{dt}=\ka[R]-\km[P].
\end{equation}   
Notice that, \eqref{eq:intra} is a first order reaction. The equilibrium would be independent of the initial concentration of the reactant.  
\subsubsection{Intermolecular reactions with two complimentary oligonucleotides}
For this class of reactions two complimentary oligonucleotides $R_1$ and $R_2$ hybridize to form the product $P$.  This class of reactions is large and includes most of the hybridization reactions involving various nucleic acids.   Hysteresis is quite commonly observed in this group of reactions.  
As for example, $R_1$ and $R_2$ may both be single stranded oligonucleotides which hybridize to produce a duplex.  \citet{bhattacharyya:et:al:2011} observed hysteresis in parallel duplex formation where one of the constituent strand was modified by LNA.  Similar examples of hysteresis in duplex formation abound.\citep{hickman:micklefield:2000,kaur:babu:maiti:2007}  
Various triplex formations where one of the reactants is a duplex and the other a single strand also show pronounced hysteresis. \citep{rougee:et:al:1992,arya:bruice:1999,maiti:2006}
 
The reaction in this case is of second order.  At equilibrium it can be expressed as:\citep{mergny:lacroix:2003,kaur:babu:maiti:2007,bhattacharyya:et:al:2011}
\begin{equation}\label{eq:rtn}
R_1+R_2\leftrightarrows P.
\end{equation}   
As before, if we assume $\ka$ and $\km$ respectively denote the rate constants of annealing and melting, 
from \eqref{eq:rtn} the rate of change of concentration of the product $P$ at time $t$ is given by
\begin{equation}\label{eq:dup}
\frac{d[P]}{dt}=\ka[R_1][R_2]-\km[P].
\end{equation}    
It is clearly seen that unlike the intramolecular reactions, the equilibrium in a bimolecular reaction depends on the initial concentration of the two reactants.  

\subsubsection{Kinetic and Thermodynamic equations in presence of hysteresis}
In both \eqref{eq:intrap} and \eqref{eq:dup} the temperature is held fixed. The concentration would vary with time, rate constants $\ka$, $\km$ would remain fixed.  If the temperature $T$ increase or decrease with time as well, the concentration as well as the rate constants will vary with it. If $dT/dt$ denotes the rate of change of $T$ at time $t$, \eqref{eq:intrap} and \eqref{eq:dup} can respectively be re-expressed as
\begin{equation}\label{eq:intrap2}
\frac{d[P]}{dT}=\left(\frac{dT}{dt}\right)^{-1}\Bigl\{\ka[R]-\km[P]\Bigr\}
\end{equation} 
and 
\begin{equation}\label{eq:dup2}
\frac{d[P]}{dT}=\left(\frac{dT}{dt}\right)^{-1}\Bigl\{\ka[R_1][R_2]-\km[P]\Bigr\}.
\end{equation}  
Here $\ka$, $\km$ and the concentrations are taken to vary with $T$.

\begin{figure}[t]
\subfigure[\label{fig:260}]{
\resizebox{2in}{1.75in}{


\setlength{\unitlength}{2cm}
\begin{picture}(6,4)(-3,-2)
\thicklines
\qbezier(0,0)(0.8853,0.8853)
(2,0.9640)
\qbezier(0,0)(-0.8853,-0.8853)
(-2,-0.9640)
\put(-2.1,-0.9640){\line(1,0){4.2}}
\put(-2.1,0.9640){\line(1,0){4.2}}
\put(-2,-1.0640){\line(0,1){2.0280}}
\put(2,-1.0640){\line(0,1){2.0280}}
\put(-2,-1.25){$T_{\mbox{min}}$}
\put(2,-1.25){$T_{\mbox{max}}$}
\put(-2.75,-.9640){$Abs_{\mbox{min}}$}
\put(-2.75,.9640){$Abs_{\mbox{max}}$}
\put(0,-.01){\vector(0,-1){.9404}}
\put(0,-.9404){\vector(0,1){.9404}}
\put(0,.01){\vector(0,1){.9404}}
\put(0,.9404){\vector(0,-1){.9404}}
\put(.1,-.4702){$FR_{\mbox{T}}$}
\put(-.45,.4702){$FP_{\mbox{T}}$}
\put(0,-.9640){\line(0,-1){.1}} 
\put(0,-1.25){$T$}
\multiput(0,0)(-.5,0){5}
{\line(-1,0){0.2}}
\put(-2.75,0){$Abs_{\mbox{T}}$}
\end{picture}

}
}
\subfigure[\label{fig:295}]{
\resizebox{2in}{1.75in}{


\setlength{\unitlength}{2cm}
\begin{picture}(6,4)(-3,-2)
\thicklines
\qbezier(0,0)(-0.8853,0.8853)
(-2,0.9640)
\qbezier(0,0)(0.8853,-0.8853)
(2,-0.9640)
\put(-2.1,-0.9640){\line(1,0){4.2}}
\put(-2.1,0.9640){\line(1,0){4.2}}
\put(-2,-1.0640){\line(0,1){2.0280}}
\put(2,-1.0640){\line(0,1){2.0280}}
\put(-2,-1.25){$T_{\mbox{min}}$}
\put(2,-1.25){$T_{\mbox{max}}$}
\put(-2.75,-.9640){$Abs_{\mbox{min}}$}
\put(-2.75,.9640){$Abs_{\mbox{max}}$}
\put(0,-.01){\vector(0,-1){.9404}}
\put(0,-.9404){\vector(0,1){.9404}}
\put(0,.01){\vector(0,1){.9404}}
\put(0,.9404){\vector(0,-1){.9404}}
\put(-.45,-.4702){$FP_{\mbox{T}}$}
\put(.1,.4702){$FR_{\mbox{T}}$}
\put(0,-.9640){\line(0,-1){.1}} 
\put(0,-1.25){$T$}
\multiput(0,0)(-.5,0){5}
{\line(-1,0){0.2}}
\put(-2.75,0){$Abs_{\mbox{T}}$}
\end{picture}

}
}
\subfigure[\label{fig:alphasch}]{
\resizebox{2in}{1.75in}{


\setlength{\unitlength}{2cm}
\begin{picture}(6,4)(-3,-2)
\thicklines
\qbezier(0,0)(-0.8853,0.8853)
(-2,0.9640)
\qbezier(0,0)(0.8853,-0.8853)
(2,-0.9640)
\put(-2.1,-0.9640){\line(1,0){4.2}}
\put(-2.1,0.9640){\line(1,0){4.2}}
\put(-2,-1.0640){\line(0,1){2.0280}}
\put(2,-1.0640){\line(0,1){2.0280}}
\put(-2,-1.25){$T_{\mbox{min}}$}
\put(2,-1.25){$T_{\mbox{max}}$}
\put(-2.4,-.9640){$0.0$}
\put(-2.4,.9640){$1.0$}
\put(0,-.9640){\line(0,-1){.1}} 
\put(0,-1.25){$T$}
\multiput(0,0)(-.5,0){5}
{\line(-1,0){0.2}}
\put(-2.75,0){$\alpha(T)$}
\end{picture}

}
}
\caption{A schematic curve of absorbance at $260$ nm \ref{fig:260} and at $295$ nm \ref{fig:295}.  Hyperchromicity is observed in the former and not in the latter.  $FR_{\mbox{T}}$ and $FP_{\mbox{T}}$ are respectively the fraction of reactant and the product at temperature $T$. 
Notice that, depending on the presence and absence of hyperchromicity, the definitions of $FR_{\mbox{T}}$ and $FP_{\mbox{T}}$ change in the plot.  \ref{fig:alphasch} A schematic diagram of $FP_{\mbox{T}}=\alpha(T)$ obtained from equations \eqref{eq:dupfrac} and \eqref{eq:dupfrac2} below.}
\label{fig:curves}
\end{figure}
 
We assume that the fraction of the product $P$, at temperature $T$, denoted by $\alpha(T)$ can be obtained from the absorbance of the heating and cooling curves at temperature $T$.   Suppose $Abs(T)$ is the absorbance at temperature $T$, $Abs_{min}$ and $Abs_{max}$ are respectively the minimum and maximum absorbance of the thermal curve under consideration.  The formula to obtain $\alpha(T)$ is determined by the presence or absence of hyperchromicity at the wavelength being used in the experiment.  
If hyperchromicity is observed, \emph{e.g.} duplex formation observed at $260$ nm, the absorbance increases with temperature.  That is higher absorbance implies higher fraction of single strands (see Figure \ref{fig:260} above).  In such cases the fraction of product at any temperature $T$ is calculated: \citep{maiti:2006,kaur:babu:maiti:2007,bhattacharyya:et:al:2011}   

\begin{equation}\label{eq:dupfrac}
\alpha(T)=\frac{Abs_{max}-Abs(T)}{Abs_{max}-Abs_{min}}.
\end{equation}  

On the other hand in the absence of hyperchromicity, e.g. triplex formation reaction recorded at $295$ nm,\citep{mergny:lacroix:2003} absorbance decreases with the increase in fraction of the single strands (see Figure \ref{fig:295}).  Thus the fraction of triplex should be calculated using the formula: \citep{maiti:2009}  

\begin{equation}\label{eq:dupfrac2}
\alpha(T)=\frac{Abs(T)-Abs_{min}}{Abs_{max}-Abs_{min}}.
\end{equation}  

Note that, by our definition for both \eqref{eq:dupfrac} and \eqref{eq:dupfrac2}, $0\le\alpha(T)\le1$, $\alpha(T_{min})=1$ which is the completely hybridized state and $\alpha(T_{max})=0$ which corresponds to the completely dissociated state.  A schematic plot of $\alpha(T)$ with temperature for both cases is shown in Figure \ref{fig:alphasch} above. 
Since the fraction decreases with temperature, the derivative is negative.  This implies that the apparent temperatures of mid-transition in the melting and annealing phases denoted by \textcolor{black}{$T_{1/2\mbox{-melt}}$ and $T_{1/2\mbox{-anneal}}$\citep{rachwalFox2007}} respectively can be approximated by the temperature corresponding to minimum value of this derivative. 

Our choice of the $Abs_{max}$ and $Abs_{min}$ as baselines, constant over the values of $T$ is a special case in a set of possible choices of baselines. Many researchers, specially Mergny and his co-authors suggest that time dependent baselines are more appropriate and strongly recommend their use throughout.\citep{mergny:lacroix:2009,mergny:lacroix:2003,mergny:lacroix:2009} 
The kinetic and thermodynamic parameters computed from the experimental data would change if the baselines are changed. In that sense, the choice of baselines are crucial.  However, as far as we could get there is no objective way of finding them. In most cases, baselines that are linear with temperature are used. 
Even then, the choice of their slopes and intercepts are subjective. In some cases (eg. \citet{rougee:et:al:1992}) baselines were obtained through separate experiments.  For intramolecular G-quadruplex formation without any evidence of hysteresis, such baselines have been obtained through model fitting. \citep{maiti:2009} 
Many authors however, prefer constant baselines. \citep{arya:bruice:1999,blasko:bruice:1996,maiti:2006,bhattacharyya:et:al:2011} Our choice is in line with them.

We should also note that, in equations \eqref{eq:dupfrac} and \eqref{eq:dupfrac2} we have made an implicit assumption that a melting curve always begin with a fully (i.e. $100\%$) folded state and end at a fully unfolded state, where as an annealing curve start and end respectively with a fully denatured and renatured states. This may not hold in many cases.  
As for example, if the temperature of midtransition is too low (eg. $20^{\circ}$ C or below), structure of the corresponding DNA may be too unstable to fold perfectly at $0^{\circ}$ C.  As a result the melting curve will not start from a $100\%$ denatured state.  On the other hand, in presence of $K^+$ ions G-Quadruplexes are too stable and a melting curve involving such DNA structures may not end at a fully denatured state.  
However, equations \eqref{eq:dupfrac} and \eqref{eq:dupfrac2} are still relevant.  If there is evidence that the nucleic acid is not fully folded at $0^{\circ}$ C or fully unfolded at $95^{\circ}$ C, 
the observed thermal curves are extrapolated beyond this range to determine the temperatures corresponding to fully renatured and denatured states.  We don't discuss the actual procedure for such extrapolation here, but equations \eqref{eq:dupfrac} and \eqref{eq:dupfrac2} can be applied to the extrapolated curves as well.

In presence of hysteresis, the concentration of the product at any temperature $T$ is different on the annealing and melting curves. \textcolor{black}{Let $\alpha_h(T)$ and $\alpha_c(T)$ respectively denote the fraction of the product at temperature $T$ at the heating (dissociation/melting) and cooling (association/annealing) curves.  Similarly we define
$\alpha_h^{(1)}(T)$ and $\alpha_c^{(1)}(T)$ respectively to be the dissociation and association rates of the product at temperature $T$.}  From equations \eqref{eq:intrap2} and \eqref{eq:dup2}, assuming equal stochiometric ratios of the reactants in \eqref{eq:dup2} and $(|dT/dt|)^{-1}=\gamma$, we get:
\begin{subequations}\label{eq:4i}
\begin{align}
\alpha_h^{(1)}(T)=\frac{d\alpha_h(T)}{dT}=&\gamma\left[\ka (1-\alpha_h(T))-\km\alpha_h(T)\right]\label{eq:4ai},\\
\alpha_c^{(1)}(T)=\frac{d\alpha_c(T)}{dT}=&-\gamma\left[\ka (1-\alpha_c(T))-\km\alpha_c(T)\right]\label{eq:4bi}.
\end{align}
\end{subequations} 
and 
\begin{subequations}\label{eq:4}
\begin{align}
\alpha_h^{(1)}(T)=\frac{d\alpha_h(T)}{dT}=&\gamma\left[\ka C(1-\alpha_h(T))^2-\km\alpha_h(T)\right]\label{eq:4a},\\
\alpha_c^{(1)}(T)=\frac{d\alpha_c(T)}{dT}=&-\gamma\left[\ka C(1-\alpha_c(T))^2-\km\alpha_c(T)\right]\label{eq:4b}.
\end{align}
\end{subequations} 

Here $C$ is the molar concentration of the reactant strands in \eqref{eq:dup2}. Notice that, in \eqref{eq:4b}, $T$ decreases with time, so $(dT/dt)^{-1}=-\gamma$. 

Equations \eqref{eq:4ai} and \eqref{eq:4bi} do not depend on the initial concentration of the reactant strand. This again indicates that the equilibrium is not dependent on the concentration factor in this case. The assumption that the concentrations of two reactant strands are equal is crucial for equations \eqref{eq:4a} and \eqref{eq:4b}. In the general case these equations would be more complex.\citep{rougee:et:al:1992, blasko:bruice:1996}
 
For any experiment $C$ and $\gamma$ are specified by the design.  Thus in order to calculate the rate constants $\ka$ and $\km$ at a given temperature $T$,  one needs to find out $\alpha_h$, $\alpha_c$, $d\alpha_h/dT$ and $d\alpha_c/dT$ at temperature $T$ and solve the equations \eqref{eq:4i} and \eqref{eq:4}.   

Once the values of $\ka$ and $\km$ are known for each temperature $T$, the thermodynamic parameters can be obtained from these values using the \emph{Arrhenius equation}.  The change in the Free Energy at the temperature $T$ (\emph{i.e}. $\Delta G^{\circ}_T$) can be directly obtained as $-RT\log(\ka/\km)$, where $\ka$ and $\km$ are the values of the rate constants at temperature $T$.  
  
The activation energies of annealing ($E_{\mbox{\scriptsize on}}$) and melting ($E_{\mbox{\scriptsize off}}$) are related to $\ka$ and $\km$ through the Arrhenius equations, 


\begin{equation}
\log(\ka)=\log(A_{\mbox{\scriptsize on}})-\frac{E_{\mbox{\scriptsize on}}}{RT},\label{eq:ka}
\end{equation}

\begin{equation}
\log(\km)=\log(A_{\mbox{\scriptsize off}})-\frac{E_{\mbox{\scriptsize off}}}{RT}\label{eq:km}.
\end{equation}

The other two thermodynamic parameters, \emph{i.e} the change in enthalpy, $\Delta H^{\circ}=E_{\mbox{\scriptsize on}}-E_{\mbox{\scriptsize off}}$ and the product of temperature and the change in entropy $T\Delta S^{\circ}=\Delta H^{\circ}-\Delta G^{\circ}_T$ can be computed once $E_{\mbox{\scriptsize on}}$ and $E_{\mbox{\scriptsize off}}$ is known.

Values of the pairs $(A_{\mbox{\scriptsize on}},E_{\mbox{\scriptsize on}})$ and $(A_{\mbox{\scriptsize off}},E_{\mbox{\scriptsize off}})$ can be determined from the \emph{Arrhenius plot} of $\log(C\ka)$ vs. $1/T$ and $\log(\km)$ vs. $1/T$ respectively. In particular, $E_{\mbox{\scriptsize on}}$ and $E_{\mbox{\scriptsize off}}$ are obtained from the slopes of the plots. We discuss this topic in more details in Section \ref{sec:rest} below.   
 
In equations \eqref{eq:4i} and \eqref{eq:4}, we have assumed that the absolute values of both the rates of cooling and heating respectively at the annealing and melting phases are $\gamma$.  That is, the rate at which the temperature goes down at the cooling phase and the rate at which it goes up at the heating phase are equal.  
Simple calculations show that in such cases, the slopes of the Arrhenius plots do not depend on $\gamma$.  This proves that the change in enthalpy, $\Delta H^{\circ}$ does not depend on $\gamma$ either.  Furthermore, even though the intercepts of the Arrhenius plots depend on $\gamma$, it can be shown that the ratio of the kinetic rate constants \emph{i.e} $\ka/\km$ is still independent of it.  
As a result $\Delta G^{\circ}_T$ and $T\Delta S^{\circ}$ are independent of $\gamma$ as well.  This implies that even if for some reason $\gamma$ is specified wrong, the kinetic analyses as described above would produce correct values of the thermodynamic parameters.  

\textcolor{black}{Let $T_m$ be the temperature of mid-transition under the hypothetical condition when the hybridisation reaction is reversible, and no hysteresis is observed.\citep{rachwalFox2007}}
We conclude this section by noting that for the intramolecular and intermolecular reactions considered here satisfy $\Delta G_{T_m}=0$ and $\Delta G_{T_m}=RT_m\log(C/2)$ respectively.\citep{mergny:lacroix:2003} Later we show how this $T_m$ can be estimated from the non-equilibrium thermal curves using our procedure.

\section{A Statistical Procedure for Data Analysis}\label{sec:stat}
\textcolor{black}{Suppose that the fraction of the product was recorded at temperatures $T_i$, $i=1,2,\ldots,n$.  For notational simplicity we denote the true fractions at $T_i$, by $\alc{i}=\alpha_c(T_i)$ and $\alh{i}=\alpha_h(T_i)$.  At any $T_i$, these fractions for both the melting and annealing curves are recorded with additive random noise.  Let the recorded (noisy) values of $\alc{i}$ and $\alh{i}$ be denoted respectively by $\alco{i}$ and $\alho{i}$.}
In order to calculate the kinetic rate constants $\ka$ and $\km$ using equations \eqref{eq:4i}, and \eqref{eq:4} one requires accurate values of the fractions and their rates of changes with the temperature from this experimental data.  Clearly, $\dalc{i}=d\alpha_c(T_i)/dT$ and $\dalh{i}=d\alpha_h(T_i)/dT$ cannot be recorded during the run of the experiment for any $T_i$.  They have to be estimated from the recorded $\alco{i}$ and $\alho{i}$ values.        

\textcolor{black}{Presence of random noise makes the estimation of $\dalc{i}$ and $\dalh{i}$ difficult.  
It is well known that the first order difference quotients e.g. \emph{i.e} $r_i=\left(\alco{(i+1)}-\alco{(i-1)}\right)/\left(T_{i+1}-T_{i-1}\right)$ can lead to extremely noisy and wrong estimates of $\dalc{i}$ (same for $\dalh{i}$).}  One possible solution used by several authors (eg. \citet{kaur:babu:maiti:2007}) is first to smooth $\alc{i}$ and then use the first order difference quotients of the smoothed data to estimate the derivative. 

Several methods of smoothing experimental data and estimating derivatives up to several orders have been discussed in Literature before. The most popular is the so called \emph{Least-squares polynomial smoothing} (LSP) introduced by \citet{savitzky:golay:1964}. 
Assuming equi-spaced data, to estimate the smoothed value of $\alc{i}$ and $\alh{i}$ this method fits polynomials to from five up to twenty five points around $T_i$.  The article provides values of the coefficients for these smoothing polynomials and for the first to fifth derivatives.  This method has been widely used computational chemistry. 
The five-point third degree LSP is used in many standard data-acquisition hardware and software packages.  

In an article \citet{Marchand:Marmet:1983} discuss several pitfalls of LSP method and propose a binomial smoothing filter as an alternative. Several other methods based on \emph{smoothing splines}, \emph{wavelets} etc. have been proposed.  
Wavelets are usually computationally demanding and the smoothing splines require one to make a crucial choice of the smoothing parameter. \citep{brabanter:et:al:2013}  In a recent article \citet{brabanter:et:al:2013} discuss a new method to estimate \textcolor{black}{$\dalc{i}$ and $\dalh{i}$} as weighted sums of difference quotients of various orders from the un-smoothed data.

An alternative to all these methods are provided by \emph{local polynomial regression} (LPR) \citep{laskorunskyi:chatterjee:dattner:2026}. This method provides an computationally efficient and easily implementable way to smooth the noisy data and estimate its derivatives.
      
\subsubsection{Local Polynomial Regression}\label{sec:lpr}
In local polynomial regression (LPR), one estimates the smoothed curve and its derivatives by fitting a polynomial within a local neighborhood of each point.  \textcolor{black}{We illustrate the procedure for the cooling curve $\alpha_c(T)$.  The procedure for the heating curve $\alpha_h(T)$ is similar.}

\textcolor{black}{Our goal is to estimate unknown function $\alpha_c$ of temperature T, from its noisy version namely $\alpha^{(obs)}_{c}$ recorded at temperatures $T_i$, $i=1,2,\ldots,n$.  More precisely the data is assumed to be generated from the model
\begin{equation}\label{eq:model}
\alpha^{(obs)}_c\left(T\right)=\alpha_c\left(T\right)+\epsilon_T,
\end{equation}
}
\textcolor{black}{and our data are the pairs $\left(T_1,\alco{1}\right)$, $\left(T_2,\alco{2}\right)$, $\ldots$, $\left(T_n,\alco{n}\right)$.}

The variable $\epsilon_T$ is the random noise, with \textcolor{black}{average i.e. the expectation} $E\left[\epsilon_T\right]=0$ and \textcolor{black}{spread i.e. the variance} $Var\left[\epsilon_T\right]=\sigma^2$ for all values of $T$.  If we assume that the $(p+1)th$ derivative of $\alpha_c(T)$ exist, we can locally approximate $\alpha_c(T)$ by a polynomial of order $p$.  For any point $T_0$ strictly with in the range of the temperature concerned and for any $T$ in a neighborhood of $T_0$, a Taylor series expansion yields, \citep{wand:jones:book,brabanter:et:al:2013}
\begin{equation}\label{eq:ex}
\alpha_c\left(T\right)\approx \alpha_c\left(T_0\right)+\alpha_c^{(1)}\left(T_0\right)\left(T-T_0\right)+\cdots+\frac{\alpha_c^{(j)}\left(T_0\right)}{j!}\left(T-T_0\right)^j+\cdots+\frac{\alpha_c^{(p)}\left(T_0\right)}{p!}\left(T-T_0\right)^p.
\end{equation} 
Here $\alpha_c^{(j)}\left(T_0\right)$ is the value of the $j$th derivative of $\alpha_c$ at $T_0$.  Note that, none of the derivatives are known and has to be estimated from the data.  

Equation \eqref{eq:ex} can be re written as:
\begin{equation}\label{eq:beta}
\alpha_c\left(T\right)\approx\sum^{p}_{j=0}\frac{\alpha_c^{(j)}\left(T_0\right)}{j!}\left(T-T_0\right)^j=\sum^p_{j=0}\beta_j\left(T-T_0\right)^j.
\end{equation}     
An estimate of the unknown vector of coefficients $\beta=(\beta_0,\beta_1,\ldots,\beta_p)$ in \eqref{eq:beta} denoted $\hat{\beta}$ is obtained as the solution to the weighted local least squares regression problem given by,
\begin{equation}\label{eq:wls}
\hat{\beta}=\mbox{arg}\min_{\beta}\sum^n_{i=1}\left\{\alco{i}-\sum^p_{j=0}\beta_j\left(T_i-T_0\right)^j\right\}^2\kr{T_i-T_0}.
\end{equation} 
We note that, the expression in \eqref{eq:ex} and \eqref{eq:beta} are local.  Thus $\hat{\beta}$ will be different for different $T_0$.

The \emph{kernel} function $K$ and the pre-specified bandwidth $h$ in \eqref{eq:wls} controls the weights put on the individual observations as well as the size of the neighborhood of $T_0$ used in estimating $\alpha_c(T_0)$.  The kernel function $K$ has to satisfy certain properties, which we don't discuss in this article  (see \citet{wand:jones:book} for details). There are several choices for $K$. Some popular ones are,
\begin{align}
K(x)&=\frac{1}{2}1_{\{|x|<1\}},&\text{(Uniform)}\label{uni}\\
K(x)&=\frac{3}{4}(1-x^2)1_{\{|x|<1\}},&\text{(Epanechnikov)}\label{epa}\\
K(x)&=(2\pi)^{-1/2}e^{-x^2/2}.&\text{(Gaussian)}\label{gau}
\end{align}   
Each Kernel puts more weight on the observations nearer to $T_0$ than those which are further away.  It may put zero weight on some observations, as in \eqref{uni} and \eqref{epa} above.  The Gaussian kernel \eqref{gau} is positive for all values of $x$ and thus takes into account all observations, but the weights of observations further away than $-3$ and $3$ are negligible.

The bandwidth $h$ plays a crucial role. Small $h$ would produce a non-smooth $m$ essentially interpolating the data.   A larger than necessary $h$ would result in over-smoothing. For $p=1$, when a straight line fitted locally, an optimal value of $h$ can be obtained from the observed data.  We refer to \citet{wand:jones:book} for details.

Once the kernel $K$ and bandwidth $h$ is specified, at any $T_0$, $\hat{\beta}$ in \eqref{eq:wls} can be easily obtained using weighted regression.  In fact, an analytic formula, involving only matrix manipulation to find $\hat{\beta}$ is available.  \textcolor{black}{From \eqref{eq:beta} the estimate $\hal{c}(T_0)$ of $\alpha_c(T_0)$ can be obtained as $\hal{c}(T_0)=\hat{\beta}_0$.}

\textcolor{black}{The same procedure can be employed to the melting curve to find estimate $\hal{h}\left(T_0\right)$ of $\alpha_h\left(T_0\right)$ for any temperature $T_0$ within the range of the data.}  


\subsubsection{Derivative Estimation}\label{sec:deriv}
Using LPR technique, any point $T_0$ the value of $\alpha_c^{(j)}(T_0)$ can be easily estimated. From \eqref{eq:beta}, the estimate is simply given by:
\begin{equation}\label{eq:mj}
\hat{\alpha}_c^{(j)}(T_0)=j!\cdot\hat{\beta}_j.
\end{equation}
\textcolor{black}{Thus the estimated value of $\dal{c}\left(T_0\right)$ and $\dal{h}\left(T_0\right)$ (denoted respectively $\hdal{c}\left(T_0\right)$, and $\hdal{h}\left(T_0\right)$) is simply the $\hat{\beta}_1$ obtained from corresponding LPR at $T_0$.}  

The estimate in \eqref{eq:mj} is in general different from the derivative of $\hal{c}$ at $T_0$.  In fact, the latter would be a bad estimate for noisy data especially for large values of $j$. \citep{brabanter:et:al:2013}   

The accuracy of $\hat{\alpha}_c^{(j)}$ depends on the value of $p$ \emph{i.e} the order of the polynomial being fit locally.  It can be shown that \citep{wand:jones:book} in order to estimate the $j$th derivative, a in some sense it is better to choose $p$ such that $p-j$ is an odd number.  

We close this section by making two comments. First of all, for LPR it is not strictly required that the temperatures $T_i$ be equi-spaced. Furthermore, $T_0$ does not need to equal $T_i$ for any $i$. It is sufficient that $T_0$ is a point within the range of observed temperatures.  \textcolor{black}{Thus we can estimate $\alpha_c$, $\alpha_h$, $\alpha_c^{(1)}$, and $\alpha_h^{(1)}$ on a temperature grid finer than the recorded one.} 
It is not also required to use that same bandwidth for the whole data.  It is natural often to use different bandwidth for different $T_0$.  Several adaptive procedures to obtain such optimal variable local bandwidths are available. 
However, LPR would not be give accurate results $T_0$ is too close to the boundary. This is an inconvenience common to many smoothing techniques.     

\subsubsection{Kinetic and Thermodynamic Parameter Estimation}\label{sec:rest}
The smoothed values of $\alc{i}$ (\emph{i.e} $\halc{i}$), $\dalc{i}$ (\emph{i.e} $\hdalc{i}$) and those of $\alh{i}$ (\emph{i.e} $\halh{i} $), $\dalh{i}$ (\emph{i.e} $\hdalh{i}$) can be determined using LPR for all values of $T_i$ from the annealing and melting curves respectively.
Once these values are available, depending on the system, equations \eqref{eq:4i} or \eqref{eq:4} can be solved at each $T_i$ to obtain the values of $C\ka$ and $\km$ at $T_i$.  Here for convenience we take $C=1$ for intramolecular reactions. 

If hysteresis is observed, the apparent half temperatures of the annealing and melting phase denoted $T_{1/2\mbox{-anneal}}$ and $T_{1/2\mbox{-melt}}$ respectively will be far apart.  They are the temperatures at which the minimum values of $\hdalc{}$ and $\hdalh{}$ were respectively obtained.    

In order to estimate the activation energies $E_{\mbox{\scriptsize on}}$ and $E_{\mbox{\scriptsize off}}$ the Arrhenius Plot is used. Because of larger change in the absorbance on the annealing and melting curves around respective $T_{1/2\mbox{-anneal}}$ and $T_{1/2\mbox{-melt}}$ $\ka$ and $\km$ can be accurately measured around these temperatures.  
Thus the Arrhenius Plot is drawn near these two temperatures.  

In order to estimate $A_{\mbox{\scriptsize on}}$ and $E_{\mbox{\scriptsize on}}$ we perform a linear regression of $\log(C\ka)$ on $1/T$ around $T_{1/2\mbox{-anneal}}$. Suppose $a_{\mbox{\scriptsize on}}$ and $b_{\mbox{\scriptsize on}}$ are respectively the estimated intercept and slope of this fit. From \eqref{eq:ka} it is clear that the estimates of $A_{\mbox{\scriptsize on}}$ (denoted $\hat{A}_{\mbox{\scriptsize on}}$) and $E_{\mbox{\scriptsize on}}$ (denoted $\hat{E}_{\mbox{\scriptsize on}}$) are given by, $\log(\hat{A}_{\mbox{\scriptsize on}})=a_{\mbox{\scriptsize on}}-\log(C)$ and $\hat{E}_{\mbox{\scriptsize on}}=-R\cdot b_{\mbox{\scriptsize on}}$.      

Similarly, suppose $a_{\mbox{\scriptsize on}}$ and $b_{\mbox{\scriptsize on}}$ are respectively the estimated intercept and slope of the regression of $\log(\km)$ with $1/T$ around $T_{1/2\mbox{-melt}}$.  Thus, from \eqref{eq:km} one can find $\log(\hat{A}_{\mbox{\scriptsize off}})=a_{\mbox{\scriptsize off}}$ and $\hat{E}_{\mbox{\scriptsize off}}=-R\cdot b_{\mbox{\scriptsize off}}$.  

We can express the change in enthalpy in terms of $b_{\mbox{\scriptsize on}}$ and $b_{\mbox{\scriptsize off}}$ as
\begin{equation}\label{eq:delh}
\Delta H^{\circ}=\hat{E}_{\mbox{\scriptsize on}}-\hat{E}_{\mbox{\scriptsize off}}=-R\left(b_{\mbox{\scriptsize on}}-b_{\mbox{\scriptsize off}}\right).
\end{equation}
Furthermore, the standard error in the computation of $\Delta H^{\circ}$, denoted $se\left(\Delta H^{\circ}\right)$ is approximately given by:
\[
se\left(\Delta H^{\circ}\right)\approx R\sqrt{\left[\left\{se\left(b_{\mbox{\scriptsize on}}\right)\right\}^2+\left\{se\left(b_{\mbox{\scriptsize off}}\right)\right\}^2\right]},
\]
where $se\left(b_{\mbox{\scriptsize on}}\right)$ and $se\left(b_{\mbox{\scriptsize off}}\right)$ are respectively the standard errors in estimating $b_{\mbox{\scriptsize on}}$ and $b_{\mbox{\scriptsize off}}$.

The change in free energy at any given temperature $T^{\star}$ is given by $\Delta G^{\circ}_{T^{\star}}=-RT^{\star}\log\left(\ka\left(T^{\star}\right)\right)+RT^{\star}\log\left(\km\left(T^{\star}\right)\right)$.  However, since the relation between $\ka$ and $\km$ with $T$ can only be accurately determined closer to $T_{1/2\mbox{-anneal}}$ and $T_{1/2\mbox{-melt}}$, \citet{bhattacharyya:et:al:2011} suggest using the predicted values obtained from the linear fits  described above.  Thus at $T^{\star}$ we get,
\begin{align}
\log\left(\ka\left(T^{\star}\right)\right)&=a_{\mbox{\scriptsize on}}-\log(C)+\frac{b_{\mbox{\scriptsize on}}}{T^{\star}},\label{eq:ats}\\
\log\left(\km\left(T^{\star}\right)\right)&=a_{\mbox{\scriptsize off}}+\frac{b_{\mbox{\scriptsize off}}}{T^{\star}},\label{eq:mts}\\
\Delta G^{\circ}_{T^{\star}}&=-RT^{\star}\left\{\left(a_{\mbox{\scriptsize on}}-\log(C)-a_{\mbox{\scriptsize off}}\right)+\frac{1}{T^{\star}}\left(b_{\mbox{\scriptsize on}}-b_{\mbox{\scriptsize off}}\right)\right\}.\label{eq:delgt}
\end{align}

From the expressions of $\Delta H^{\circ}$ and $\Delta G^{\circ}_{T^{\star}}$ above, we can express $T^{\star}\Delta S^{\circ}=\Delta H^{\circ}-\Delta G^{\circ}_{T^{\star}}$ in terms of $a_{\mbox{\scriptsize on}}$, $a_{\mbox{\scriptsize off}}$, $b_{\mbox{\scriptsize on}}$ and $b_{\mbox{\scriptsize off}}$.  The explicit expression turns out to be,
\begin{equation}\label{eq:tdels}
T^{\star}\Delta S^{\circ}=RT^{\star}\left(a_{\mbox{\scriptsize on}}-\log(C)-a_{\mbox{\scriptsize off}}\right).
\end{equation}

We approximate the standard error in estimating $T^{\star}\Delta S^{\circ}$, denoted by $se\left(T^{\star}\Delta S^{\circ}\right)$ as:
\[
se\left(T^{\star}\Delta S^{\circ}\right)\approx RT^{\star}\sqrt{\left[\left\{se\left(a_{\mbox{\scriptsize on}}\right)\right\}^2+\left\{se\left(a_{\mbox{\scriptsize off}}\right)\right\}^2\right]},
\]
where $se\left(a_{\mbox{\scriptsize on}}\right)$ and $se\left(a_{\mbox{\scriptsize off}}\right)$ are the standard errors in estimating $a_{\mbox{\scriptsize on}}$ and $a_{\mbox{\scriptsize off}}$ respectively. 

Finally, we the error in estimation of $\Delta G^{\circ}_{T^{\star}}$, i.e. $se\left(\Delta G^{\circ}_{T^{\star}}\right)$ can be approximated as:
\[
se\left(\Delta G^{\circ}_{T^{\star}}\right)\approx RT^{\star}\sqrt{\left[\left\{se\left(a_{\mbox{\scriptsize on}}\right)\right\}^2+\left\{se\left(a_{\mbox{\scriptsize off}}\right)\right\}^2+\frac{1}{(T^{\star})^2}\left[\left\{se\left(b_{\mbox{\scriptsize on}}\right)\right\}^2+\left\{se\left(b_{\mbox{\scriptsize off}}\right)\right\}^2\right]\right]}
\]

Clearly, $\Delta H^{\circ}$ and $T^{\star}\Delta S^{\circ}$ can be obtained directly from the linear fits on the Arrhenius plot.  $\Delta G^{\circ}_{T^{\star}}$ can then be calculated from $\Delta H^{\circ}$ and $T^{\star}\Delta S^{\circ}$.

\subsubsection{Finding the temperature of mid-transition}
\textcolor{black}{The temperature of mid-transition for the reversible reaction under idealised condition,} $T_m$ can be computed by solving \eqref{eq:delgt} with known values of $\Delta G_{T_m}$.  For intramolecular reactions it is known that $\Delta G_{T_m}=0$, which implies
\begin{equation}
T_m=\frac{b_{\mbox{\scriptsize off}}-b_{\mbox{\scriptsize on}}}{a_{\mbox{\scriptsize on}}-a_{\mbox{\scriptsize off}}}\text{ $~^{\circ}$ K}.
\end{equation}
For intermolecular reactions with two complimentary oligonucleotides $\Delta G_{T_m}=RT_m\log(C/2)$.  This leads to
\begin{equation}
T_m=\frac{b_{\mbox{\scriptsize off}}-b_{\mbox{\scriptsize on}}}{a_{\mbox{\scriptsize on}}-a_{\mbox{\scriptsize off}}-\log(2)}\text{ $~^{\circ}$ K}.
\end{equation}
In both cases $T_m$ does not depend on either $\gamma$ or $C$.
\section{Software Implementation}
The whole procedure described above can be implemented with available statistical softwares.  We provide a free web-based application called ``Anhysnuc'' for this purpose. The application provides a one-step solution to the problem in real time requiring very little tuning. From an input of an absorbtion spectra, this web based application
  produces full analyses of the curves including determination of kinetic and thermodynamic parameters. The numerical results, including the figures can be easily downloaded for further use. Anhysnuc uses packages from a free statistical repository called {\tt R}. \citep{R_manual} The web-based software can be found at \href{https://sanjaychaudhuri.shinyapps.io/anhysnuc/}{https://sanjaychaudhuri.shinyapps.io/anhysnuc/}.

As we have mentioned before, determination of the bandwidth $h$ used in LPR is crucial.  A bad choice of $h$ may produce a over or under smoothed solution, which will affect the values of $\ka$, $\km$ and other parameters at a later stage.  We are not required to use the same $h$ for all the points.  
The thermal curves do not change much near the two ends, but large changes in absorbance are observed around $T_{1/2\mbox{-anneal}}$ and $T_{1/2\mbox{-melt}}$. Thus it is advisable to use different bandwidths for different parts of the curves.

The optimal bandwidth depends on the choice of kernel function $K$ as well.  There is a vast literature in statistics dealing with determination of optimal global or local optimal bandwidths. We skip the technical details here.

The web-based software Anhysnuc uses the a function named {\tt lokerns} from the package named {\tt lokern} \citep{lokern} in {\tt R}.  In order to find the smoothed values $\halc{i}$ of $\alc{i}$ at any temperature $T_i$, the {\tt lokerns} function uses Gaussian kernel to compute an adaptive local plug-in bandwidth. 
With this variable local bandwidth and using Gaussian kernel the values of $\halc{i}$ are finally calculated.  The smoothed values $\halh{i}$ and derivatives $\hdalc{i}$ and $\hdalh{i}$ at each $T_i$ are calculated similarly.  We use separate runs of {\tt lokerns} to calculate the above four functions.  Thus the optimal local bandwidths for each of them are different.  
After obtaining $\hdalc{}$ and $\hdalh{}$, The values of $T_{1/2\mbox{-anneal}}$ and $T_{1/2\mbox{-melt}}$ can be easily obtained.  One however, needs to be careful about the spurious low values of the derivatives near the ends of the temperature range.

From the available values of $\halc{i}$, $\halh{i}$, $\hdalc{i}$ and $\hdalh{i}$ at each $T_i$, it is easy to solve equations \eqref{eq:4i} and \eqref{eq:4} to obtain the values of $C\ka$ and $\km$. For the purpose of the Arrhenius plot it is sufficient to obtain these values near $T_{1/2\mbox{-anneal}}$ and$T_{1/2\mbox{-melt}}$.  However, we compute $C\ka$ and $\km$ for all available temperatures.  This is computationally wasteful but only slightly.     
                
The linear fits in the Arrhenius plots are automatically obtained from the web-based app. First it estimates the values of $a_{\mbox{\scriptsize on}}$, $b_{\mbox{\scriptsize on}}$, $a_{\mbox{\scriptsize off}}$ and $b_{\mbox{\scriptsize off}}$, and their standard errors. The thermodynamic parameters at a temperature $T^{\star}$ and their standard errors are calculated from these using the formula in equations \eqref{eq:delh}, \eqref{eq:delgt}, \eqref{eq:tdels}, etc.  
    
\section{Illustrative Examples}

\subsection{Application to Real Datasets}
In this section we present two illustrative examples of the procedure described above.  Our first example considers intramolecular G-Quadruplex formation of two specific RNA strands. In the second example we consider duplex formation by LNA modified DNA strands at two different pH values.    
\subsubsection{Intramolecular G-Quadruplex formation}
\citet{pandey:maiti:2013} considered the effect of loops and G-quartets on the stability of RNA G-quadruplexes.  We select two RNA strands from this article.  They are namely, BAP1 ($5^{\prime}$-GGGUGGGCCCUGGGC UCUGGG-$3^{\prime}$) and CCDC64 ($5^{\prime}-$GGGCCCCAUGGGUCCGGGAGGG$-3^{\prime}$).  For both of these strands hysteresis was observed for a temperature gradient of $0.25^{\circ}$C per minute.

\begin{figure}[ht]
\subfigure[Smoothed data.\label{fig:l:bap1}]{
\resizebox{3in}{!}{\includegraphics{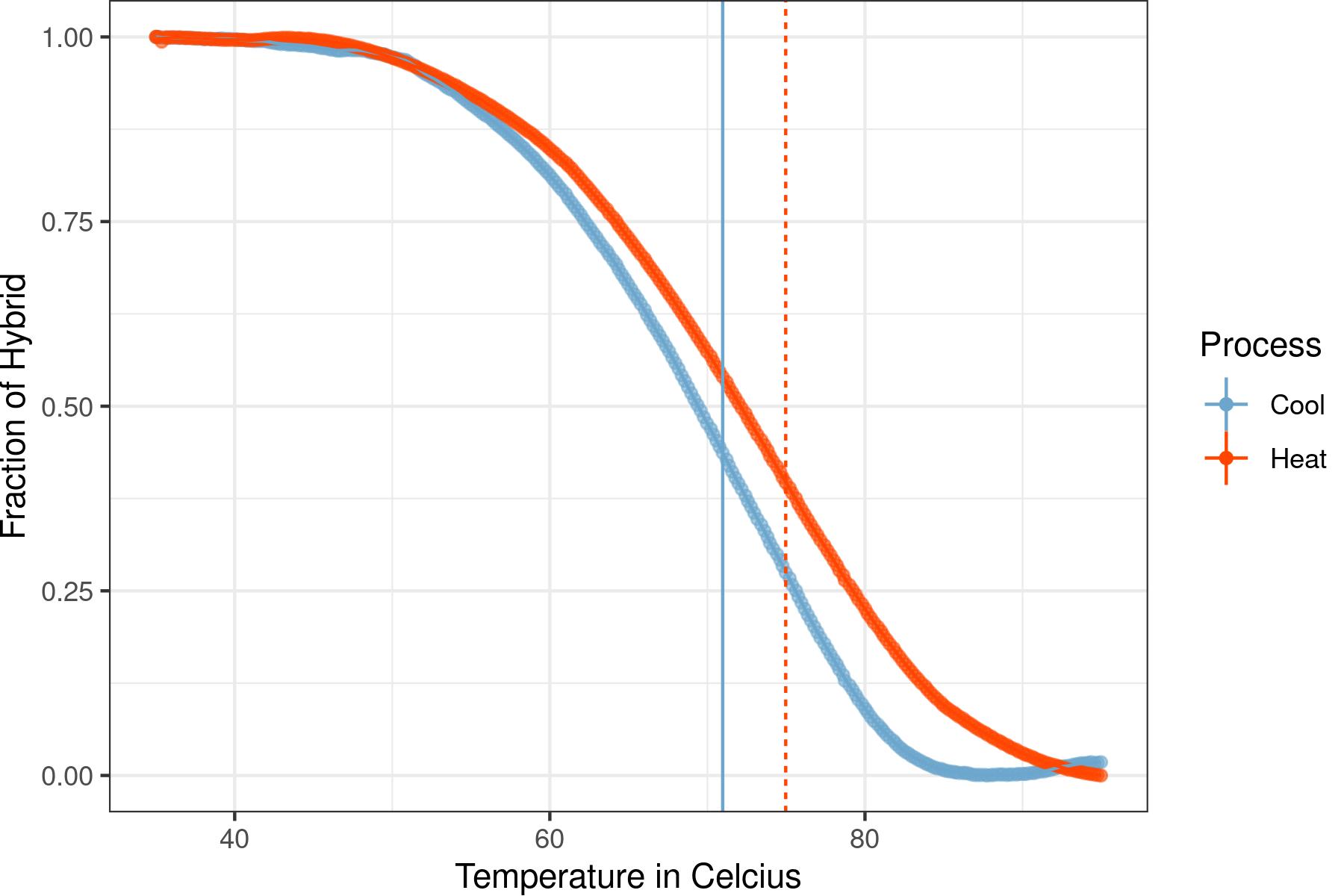}}
}
\subfigure[Derivative.\label{fig:d:bap1}]{
\resizebox{3in}{!}{\includegraphics{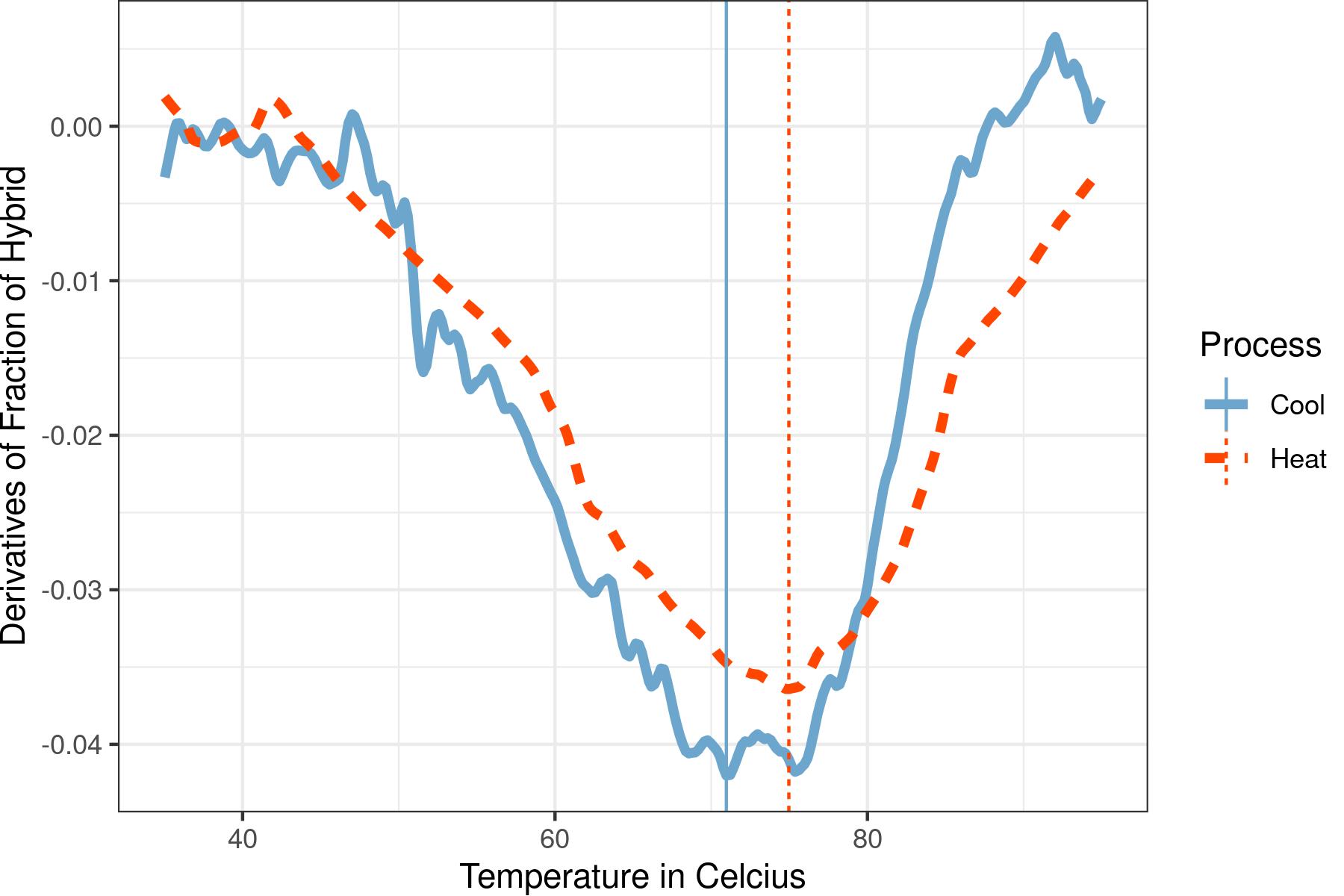}}
}
\subfigure[Arrhenius Plot.\label{fig:ah:bap1}]{
\resizebox{3.2in}{!}{\includegraphics{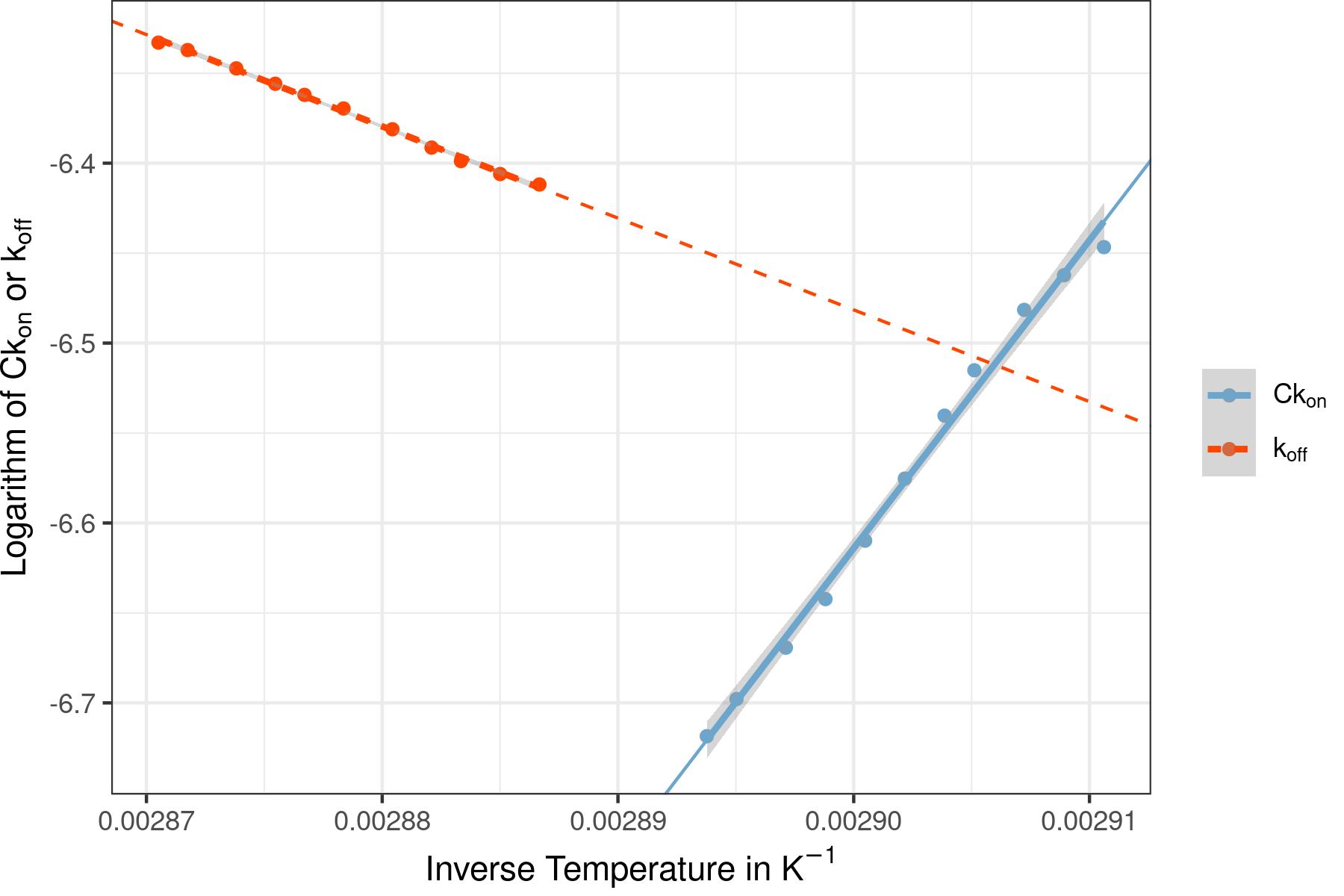}}
}
\caption{Smoothed curve \ref{fig:l:bap1}, Derivative \ref{fig:d:bap1} and Arrhenius Plot \ref{fig:ah:bap1} for the intramolecular G-quadruplex formation described in \citet{pandey:maiti:2013}. The fraction of the Quadruplex was calculated using Equation \eqref{eq:dupfrac}.  The results were obtained from the proposed software.  The red lines correspond to melting and the blue lines correspond to annealing.
} 
\label{fig:bap1}
\end{figure}

\begin{table}[b]
\caption{Kinetic and thermodynamic parameters of the hybridization experiment of intramolecular RNA G-Quadruplexes formation from \citet{pandey:maiti:2013}.  The quoted values of $\ka$, $\km$, $T\Delta S^{\circ}$ and $\Delta G^{\circ}_T$ are those predicted at $25.0^{\circ}$ C by the linear fits of the Arrhenius plot.}
\medskip
\begin{tabular}{|l|c|c|c|c|c|c|c|}\hline\hline
Strand&$T_{1/2\mbox{-anneal}}$&$T_{1/2\mbox{-melt}}$&$\ka$&$\km$&$\Delta H^{\circ}$&$T\Delta S^{\circ}$&$\Delta G^{\circ}_T$\\
&&&{\tiny$(M^{-1}s^{-1})$}&{\tiny$(s^{-1})$}&{\tiny$(\mbox{kcal mol}^{-1})$}&{\tiny$(\mbox{kcal mol}^{-1})$}&\tiny{$(\mbox{kcal mol}^{-1})$}\\\hline
BAP1&$71.0^{\circ}$C&$75.0^{\circ}$C&$1.15$&$3.11\times 10^{-4}$&\thead{$-36.7$\\$\pm 1.54$}&\thead{$-31.8$\\$\pm 1.33$}&\thead{$-4.89$\\$\pm 2.03$}\\\hline
CCDC64&$68.6^{\circ}$C&$72.2^{\circ}$C&$13.1$&$2.33\times 10^{-5}$&\thead{$-62.0$\\$\pm 6.59$}&\thead{$-54.1$\\$\pm 5.74$}&\thead{$-7.89$\\$\pm 8.74$}\\\hline
\end{tabular}
\label{tab:quad}
\end{table}

The topic of quadruplex melting has been studied by many authors before.  We refer to \citet{rachwalEtAl2007,rachwalFox2007,rachwalBrownFox2007,sekiboFox2017} for more details.  The kinetics of DNA quadruplex formation has been studied in \citet{rachwalEtAl2007}.

The fitted values, derivative and the Arrhenius Plot for the BAP1 strand is presented in Figure \ref{fig:bap1}.  The figures were drawn using the basic plotting function in {\tt R}.  The fitted values (Figure \ref{fig:l:bap1}) show pronounced hysteresis which is reflected in the difference in the values of $T_{1/2\mbox{-melt}}$ and $T_{1/2\mbox{-anneal}}$ as observed in Figure \ref{fig:d:bap1}.  The Arrhenius plot is in Figure \ref{fig:ah:bap1}.

In Table \ref{tab:quad} we present the values of the kinetic and thermodynamic parameters of the G-quadruplex formation of BAP1 and CCDC64 obtained using our method. The quoted values of $\ka$ and $\km$ are for $25.0^{\circ}$C. These values were found by extrapolating the linear fits of their corresponding values near $T_{1/2\mbox{-anneal}}$ and $T_{1/2\mbox{-melt}}$ in the Arrhenius Plot (see equations \eqref{eq:ats} and \eqref{eq:mts}).
The values of $T_m$ turns out to be $71.0^{\circ}$C and $68.5^{\circ}$C respectively.  These values are pretty close to corresponding temperatures reported by \citet{pandey:maiti:2013}, which are $73.0^{\circ}$C and $70.0^{\circ}$C for BAP1 and CCDC64 respectively.  This shows that using the proposed procedure one can compute the value of $T_m$ quite accurately from non-equilibrium thermal curves. 

\subsubsection{LNA modified duplex formation}

For our second illustrative example, we consider data from a hybridization experiment of duplex formation consisting of PuP ($5^{\prime}-\mbox{AGAAAGAGAAGA}-3^{\prime}$) and Py0 ($5^{\prime}-\mbox{TCTTTCTCTTCT}-3^{\prime}$).  These sequences were used by \citet{bhattacharyya:et:al:2011}, who further introduced LNA modification at the cytosine bases of Py0.  

They performed the melting and annealing both with a temperature gradient of $0.5^{\circ}$C per minute with $C=10^{-6}$ mol and reported marked hysteresis in the hybridization of PuP and LNA modified Py0.  Extent of this hysteresis increased with the number of LNA modifications.
We choose data from the hybridization of Py0 and Py4 ($5^{\prime}-\mbox{T}$$\mbox{C}^{\mbox {L}}$$\mbox{TTT}$$\mbox{C}^{\mbox {L}}$$\mbox{T}$$\mbox{C}^{\mbox {L}}$$\mbox{TT}$$\mbox{C}^{\mbox {L}}$$\mbox{T}-3^{\prime}$), with LNA modification (denoted by $\mbox{C}^{\mbox {L}}$) on all four cytosine bases of Py0 for our purpose.

\begin{figure}[h!]
\subfigure[Smoothed data.\label{fig:l:p:ph7:3}]{
\resizebox{3in}{!}{\includegraphics{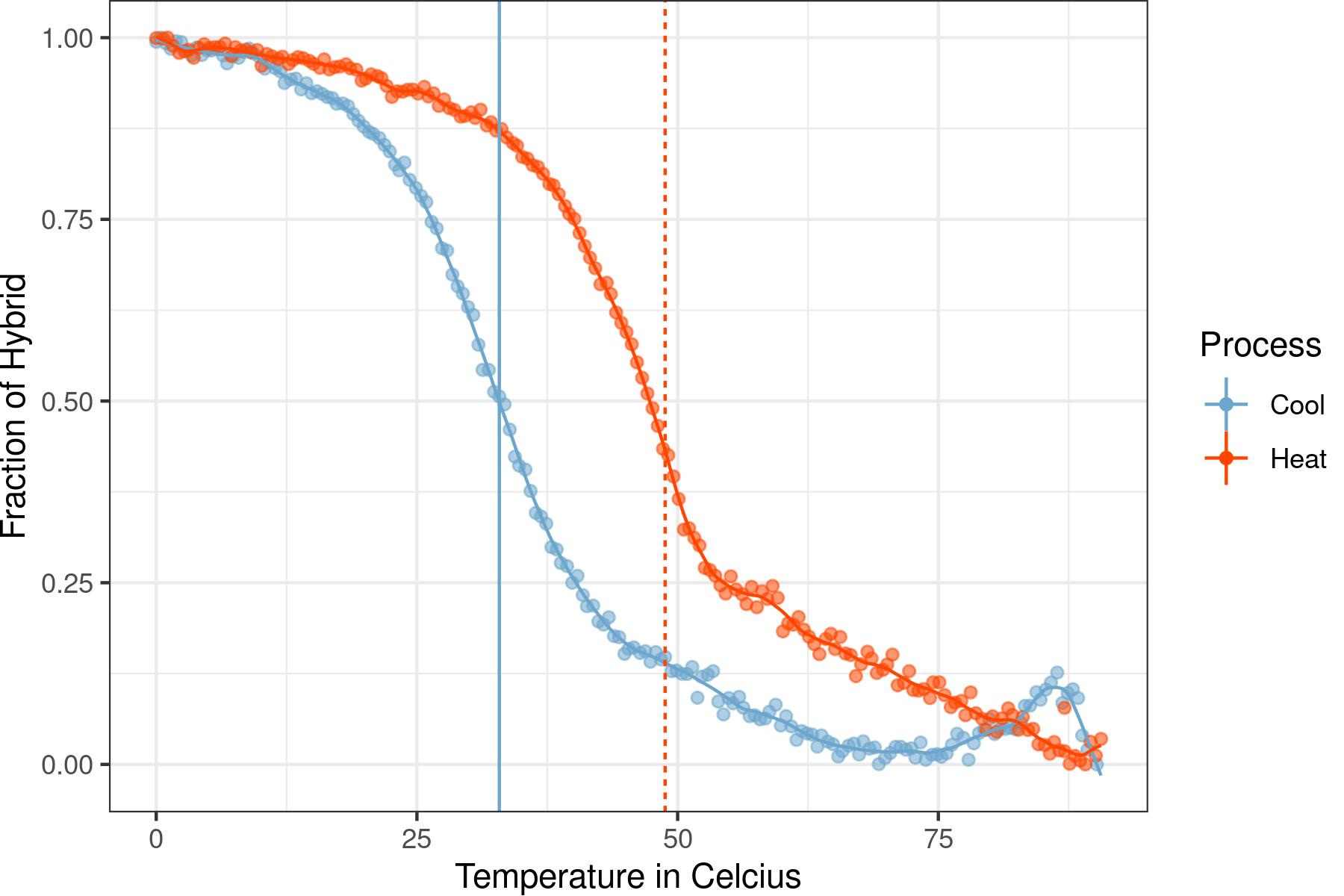}}
}
\subfigure[Derivative.\label{fig:d:p:ph7:3}]{
\resizebox{3in}{!}{\includegraphics{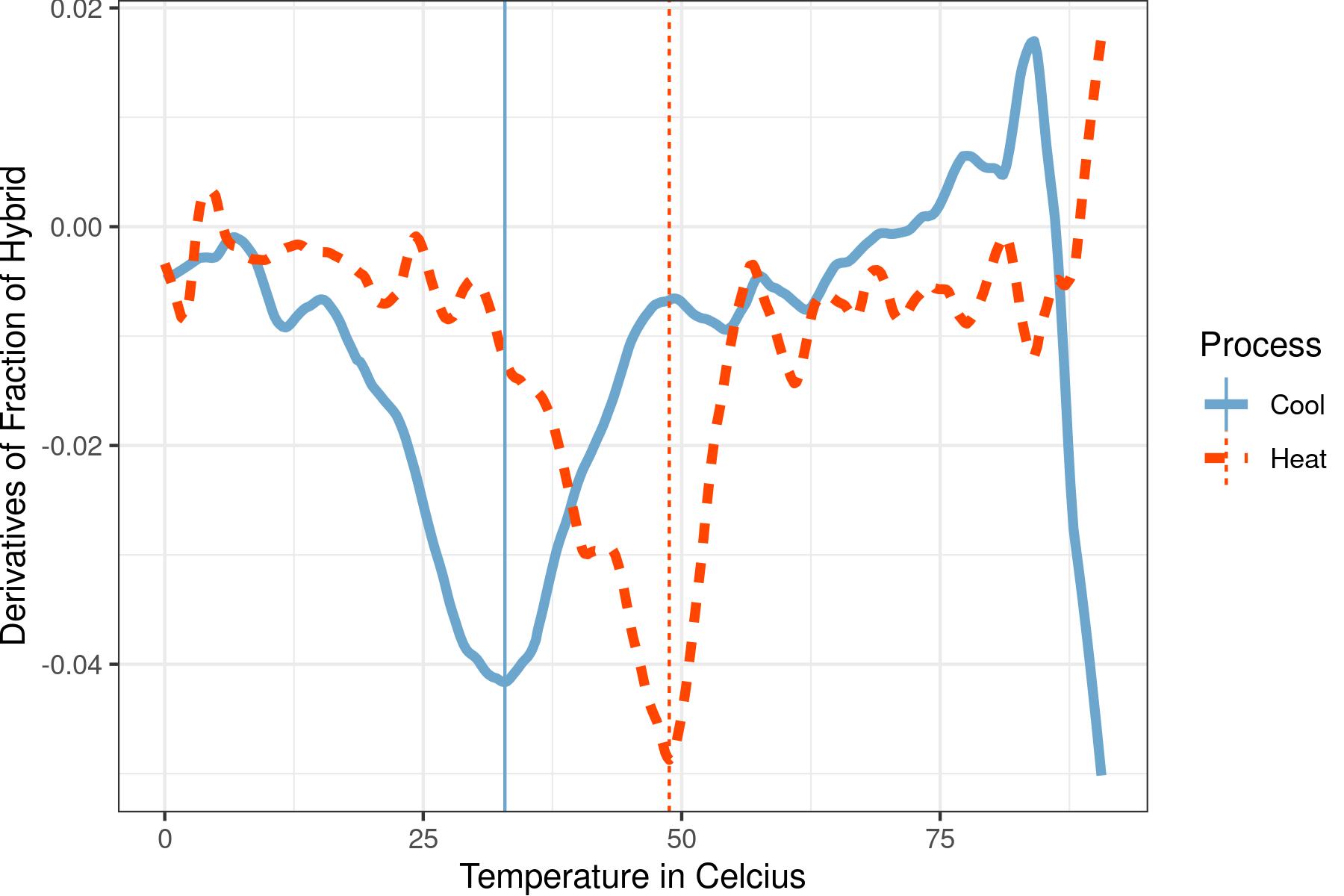}}
}
\subfigure[Arrhenius Plot.\label{fig:ah:p:ph7:3}]{
\resizebox{3.2in}{!}{\includegraphics{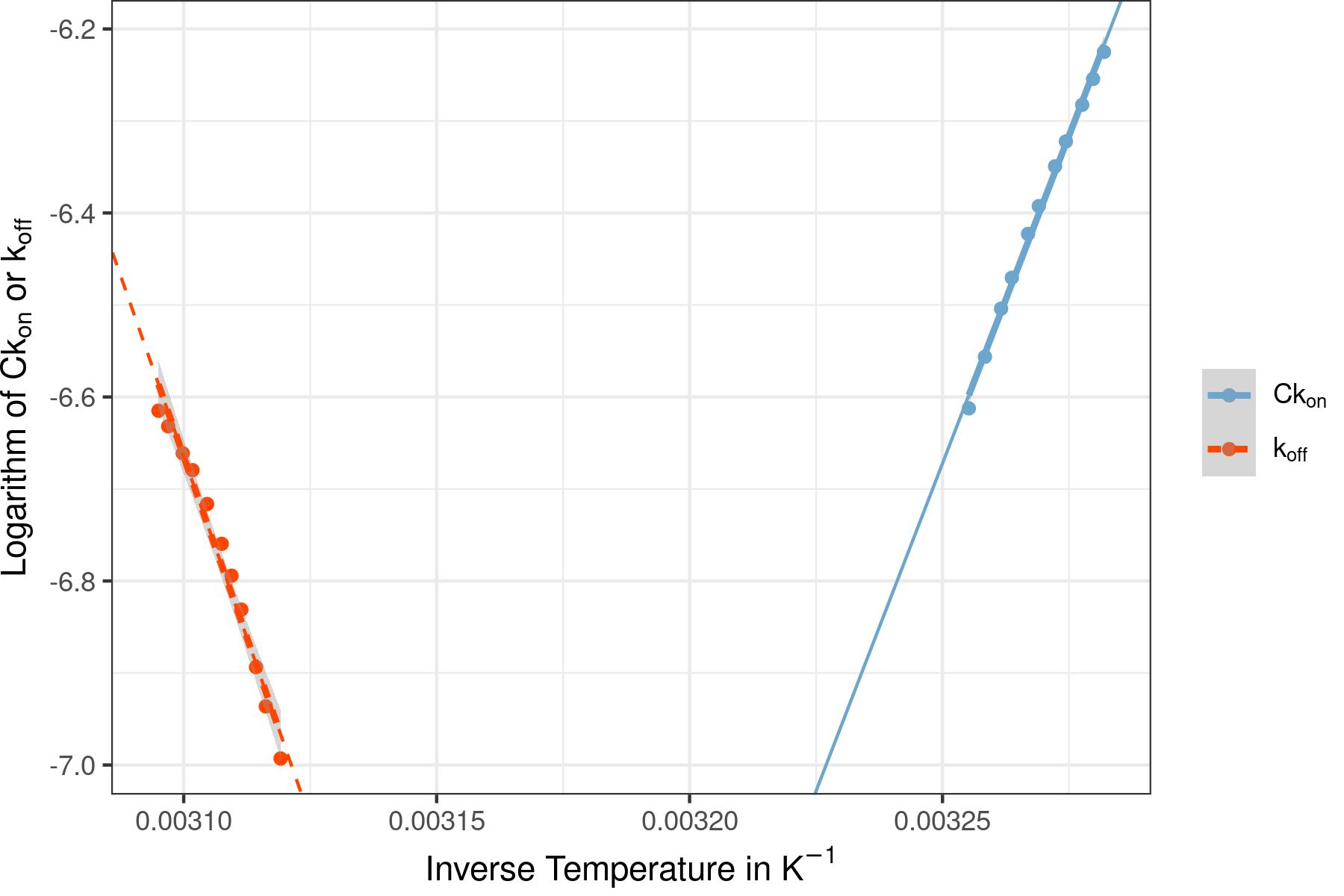}}
}
\caption{Smoothed curve \ref{fig:l:p:ph7:3}, Derivative \ref{fig:d:p:ph7:3} and Arrhenius Plot \ref{fig:ah:p:ph7:3} for PuP/Py4 sequence as described in \citet{bhattacharyya:et:al:2011}. The red lines correspond to melting and the blue lines correspond to annealing.  The fraction of the duplex was calculated using Equation \eqref{eq:dupfrac}.  The results were obtained from the proposed software.  Note that our value for $T_{1/2\mbox{-melt}}$ and $T_{1/2\mbox{-anneal}}$ are close to the values of these parameters reported in \citet{bhattacharyya:et:al:2011}.} 
\label{fig:py4}
\end{figure}

\begin{table}[th]
\caption{Kinetic and thermodynamic parameters of the hybridization experiment of PuP/Py0 and PuP/Py4 reported in \citet{bhattacharyya:et:al:2011}.  The quoted values of $\ka$, $\km$, $\Delta H^{\circ}$, $T\Delta S^{\circ}$ and $\Delta G^{\circ}_T$ are those predicted at $25.0^{\circ}$C by the linear fits of the Arrhenius plot.}
\medskip
\begin{tabular}{|l|c|c|c|c|c|c|c|c|}\hline\hline
Strand&pH&$T_{1/2\mbox{-anneal}}$&$T_{1/2\mbox{-melt}}$&$\ka$&$\km$&$\Delta H^{\circ}$&$T\Delta S^{\circ}$&$\Delta G^{\circ}_T$\\
&&&&{\tiny$(M^{-1}s^{-1})$}&{\tiny$(s^{-1})$}&{\tiny$(\mbox{kcal mol}^{-1})$}&{\tiny$(\mbox{kcal mol}^{-1})$}&\tiny{$(\mbox{kcal mol}^{-1})$}\\\hline
Py0&$5.0$&$25.0^{\circ}$C&$28.0^{\circ}$C&$3.70\times 10^{3}$&$8.88\times 10^{-4}$&\thead{$-53.8$\\$\pm 1.26$}&\thead{$-44.8$\\$\pm 1.25$}&\thead{$-9.1$\\$\pm 1.77$}\\\cline{2-9}
&$7.0$&$15.9^{\circ}$C&$19.2^{\circ}$C&$8.90\times10^{2}$&$8.96\times 10^{-4}$&\thead{$-41.8$\\$\pm 2.43$}&\thead{$-33.6$\\$\pm 2.51$}&\thead{$-8.2$\\$\pm 3.49$}\\\hline
Py4&$5.0$&$53.0^{\circ}$C&$71.0^{\circ}$C&$7.65\times 10^{5}$&negligible&\thead{$-109.8$\\$\pm 2.50$}&\thead{$-89.2$\\$\pm 2.19$}&\thead{$-20.6$\\$\pm 3.32$}\\\cline{2-9}
&$7.0$&$32.9^{\circ}$C&$48.8^{\circ}$C&$5.70\times 10^3$&$2.27\times 10^{-5}$&\thead{$-60.0$\\$\pm1.70 $}&\thead{$-48.5$\\$\pm 1.59$}&\thead{$-11.5$\\$\pm 2.33$}\\\hline
\end{tabular}
\label{tab:res}
\end{table}

The hybridization of PuP/Py4 shows a marked Hysteresis for both pH ($5.0$ and $7.0$) values. We present the fitted values, derivative and the Arrhenius Plot for pH $7.0$ in Figure \ref{fig:py4}.  In Table \ref{tab:res} the values of the kinetic and thermodynamic parameters of the hybridization experiment of PuP/Py0 and PuP/Py4 obtained using our method are presented.  As before the quoted values of $\ka$ and $\km$ are for $25.0^{\circ}$ C and were obtained by extrapolating the linear fits in the Arrhenius Plot as described above.




\subsection{Validatory Simulation Analysis}
In this section we apply the proposed procedure on simulated data generated to mimic experimental datasets.  

The data was generated by numerically solving the pair of differential equations for the intermolecular reaction (10a) and (10b) using the RK4, i.e., Runge-Kutta order $4$ method in the temperature range $0^o$C to $100^o$C at a time step of $0.05^o$C \citep{paul:peng:burman:2011}.  To replicate the form of the curves, it is required that both $\ka$ and $\km$ need to be temperature-dependent.  This is ensured by specifying $\log(\ka)$ and $\log(\km)$ follow the Arrhenius equations (see \eqref{eq:ka} and \eqref{eq:km}).  Unlike \citet{zhaoEtAl2004}, we don’t use fixed $\ka$ and $\km$.  The parameter values were fixed as follows: $\gamma=600/5$, $A_{on}=1.996734\times10^{-24}$, $E_{on}=-29.45961$, $A_{off}=1.071917\times10^{17}$, $E_{off}=29.63646$, $R=1/500$, $C=10^{-6}$.
These parameter values correspond to those of PuP/Py4 sequence at Ph$=7.0$ and temperature $25^oC$. It should be noted that the curves are sensitive to the values of the parameters. An arbitrary choice would not produce sigmoid curves with pronounced hysteresis.

Random noise was then added to the absorbance values at each temperature. The noisy data were processed by the software, and the estimated values of $\ka$ and $\km$ at $25^o$C were compared with their true values ($\ka= 5848.012$, $\km= 2.720231×10^{-5}$)\footnote{The numbers differ slightly from Table \ref{tab:res} becuase of rounding in the input parameters.} at the same temperature. The results presented below are based on $1000$ replications of the procedure described above.

\subsubsection{Symmetric error:} We use a Gaussian error with zero mean and standard deviations varying as $0.0025$, $0.005$, $0.01$, and $0.025$.  It was found that a higher error standard deviation led to an unrealistically low signal-to-noise ratio in the data (see Figure \ref{fig:curves}). For comparison, the standard error obtained from smoothing the data was approximately $0.01$.

\begin{figure}[t]
  \resizebox{3.5in}{!}{\includegraphics{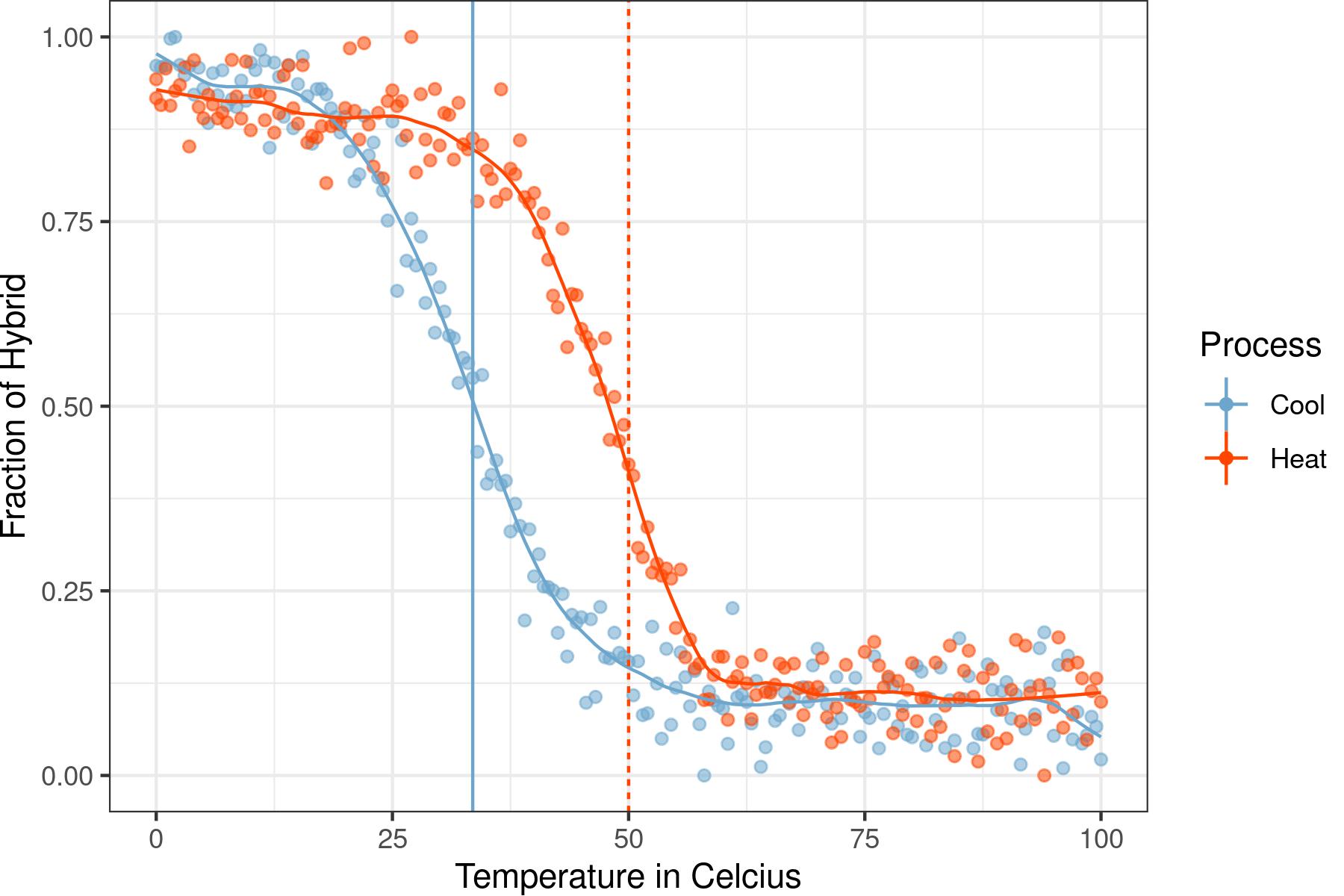}}
  \caption{An example of a simulated noisy melting and annealing curve with a noise standard deviation of $0.05$.  The low signal-to-noise ratio is evident, especially near $25^o$C.}
  \label{fig:curve}
  \end{figure}

The bias (i.e. the mean difference between the estimated and the true value), relative bias (bias/true value), the $.025th$, and the $.975th$ percentile of the estimates can be found in Table \ref{tab:sym}. Since the distribution of the estimates are skewed, the percentiles are better measures of spread than the standard errors or the root mean squared errors.

\begin{table}[h]
  \caption{The raw and relative bias, estimate-percentiles in the estimation of $\ka$ and $\km$ for simulated Gaussian noisy data with different standard deviations from the intermolecular hybridization experiment.}
  \label{tab:sym}
  \begin{tabular}{|l|r|r|c|r|r|c|}\hline\hline
  \thead{Standard\\ deviation of\\ Gaussian\\ Error}&	\multicolumn{3}{c|}{\thead{Errors in Estimation of $\ka$\\ True Value $5848.012$}}&\multicolumn{3}{c|}{\thead{Errors in Estimation of $\km$\\ True Value $2.72×10^{-5}$}}\\\hline
&bias&\thead{Relative\\bias}&\thead{Percentiles\\(.025,.975)}&\thead{Bias\\$\times10^{-6}$}&\thead{Relative\\bias}&\thead{Percentiles\\$\times10^{-5}$\\(.025,.975)}\\\hline
0.0025&	11.64&	0.2\%&	(5365.93,6363.57)&	0.83&	3.4\%&	(2.22,3.45)\\
0.005&	-59.08&	-1.0\%&	(5048.11,6569.93)&	1.15&	4.2\%&	(1.75,4.04)\\
0.010&	-232.76&	-4.0\%&	(4584.77,6810.25)&	1.92&	7.1\%&	(1.31,4.82)\\
0.025&	-727.48&	-12.4\%&	(3608.32,7120.66)&	3.85&	14.2\%&	(0.73,6.27)\\\hline
\end{tabular}
  \end{table}
  
\begin{figure}[h]
\subfigure[\label{fig:sym:kon}]{
\resizebox{3.1in}{!}{\includegraphics{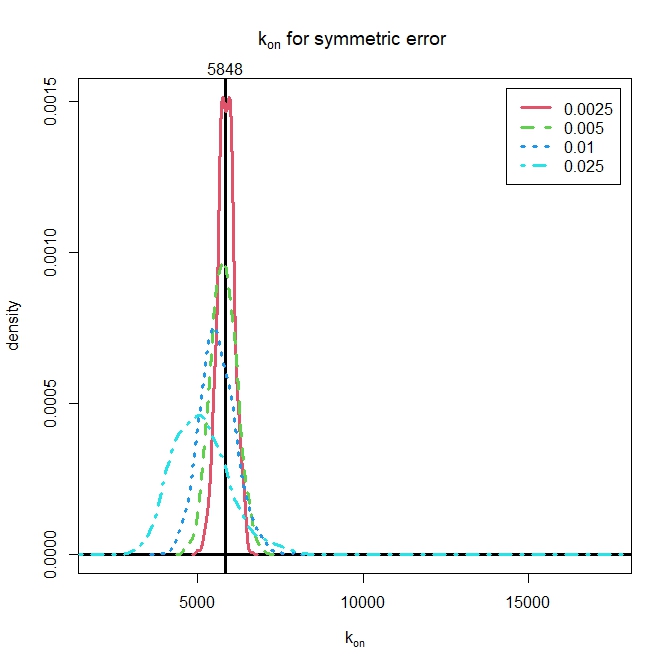}}
}
\subfigure[\label{fig:sym:koff}]{
\resizebox{3.1in}{!}{\includegraphics{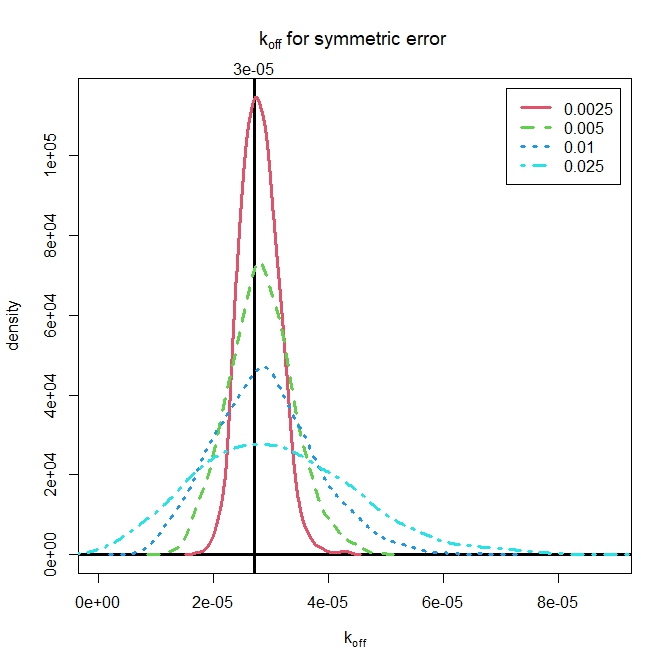}}
}
\caption{Density plots of the estimated $\ka$ and $\km$ from the $1000$ replications of the simulated noisy melting and annealing curves. The standard deviation of the Gaussian noise varies. 
} 
\label{fig:kSymErr}
\end{figure}

The density plots of the estimated $\ka$ and $\km$ obtained from the $1000$ replications can be found in Figure \ref{fig:kSymErr}.  The density plots are skewed, with long tails to the right. From the density plots, it seems that as the standard deviation of the error increases, i.e. signal-to-noise ratio decreases, on average, $\ka$ is underestimated. However, if the standard deviation of the error is below $0.01$, such biases are small (below $5\%$). 

The density plots for $\km$ turn out to be more symmetric, and their spread increases with the loss of the signal-to-noise ratio.  The relative bias is positive, but remains below $7\%$ for error standard deviation below $0.01$.

\subsubsection{Asymmetric error:} 
\begin{figure}[t]
  \resizebox{3.5in}{!}{\includegraphics{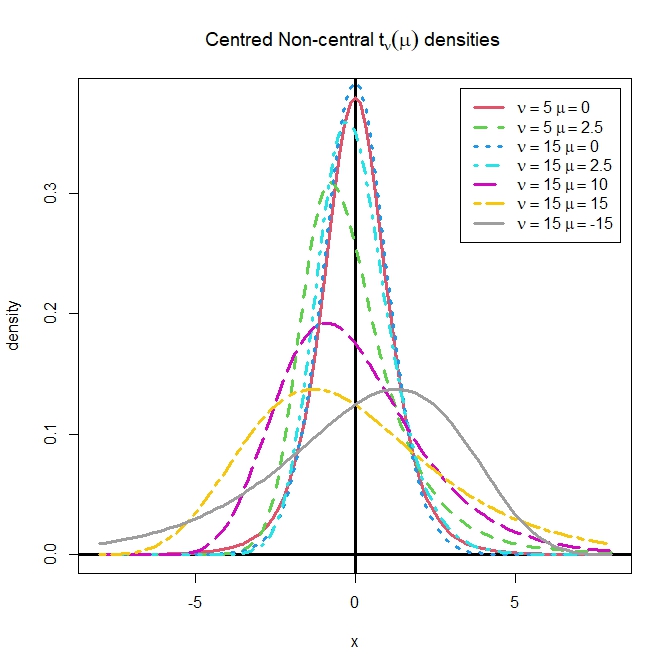}}
  \caption{Density plots of non-central $t_{\nu}(\mu)$ distribution for various values of $\nu$ and $\mu$.}
  \label{fig:nCt}
  \end{figure}

To generate data with asymmetric error, we use a centred and scaled non-central $t_{\nu}(\mu)$ distribution, with different values of $\nu$ and $\mu$.  The non-central $t_{\nu}(\mu)$ distribution is skewed, with non-zero mean and variance depending on $\nu$ and $\mu$. It is well known that the $t$ distribution has thicker tails than the Gaussian distribution and, therefore, is expected to generate more extreme error values. The non-centrality creates skewness, which, if positive, would result in larger positive errors than negative errors. If $\mu=0$, the distribution is symmetric. Additionally, for large values of $\nu$, the tails of the distribution decrease to the tails of a Gaussian random variable. Even though we centred the distribution to have a zero mean, it is still skewed. Density plots of some $t_{\nu}(\mu)$ densities can be found in Figure \ref{fig:nCt}.

 In order to use this distribution as an error distribution, the mean was subtracted from the simulated errors. Finally, the errors were multiplied by a factor of $0.005$ in order to keep the standard deviation below or near $0.01$.

\begin{table}[h]
  \caption{The raw and relative bias, estimate-percentiles in the estimation of $\ka$ and $\km$ for simulated centered and scaled non-central t noisy data with different standard deviations from the intermolecular hybridization experiment.}
  \label{tab:asym}
  \begin{tabular}{|l|r|r|c|r|r|c|}\hline\hline
\thead{Standard deviation\\$(\nu,\mu)$}&\multicolumn{3}{c|}{\thead{Errors in Estimation of $\ka$\\True Value $5848.012$}}&\multicolumn{3}{c|}{\thead{Errors in Estimation of $\km$\\True Value $2.72\times10^{-5}$}}\\\hline
&Bias&\thead{Relative\\bias}&\thead{Percentiles\\(.025,.975)}&\thead{Bias\\$\times10^{-6}$}&\thead{Relative\\bias}&\thead{Percentiles\\$\times10^{-5}$\\(.025,.975)}\\\hline
0.006 (5,0)&	-95.15&	1.6\%&	(4848.53,6726.63)&	1.15&	4.2\%&	(1.54,4.31)\\
0.015 (5,2.5)&	-344.51&-5.9\%&	(4338.81,6906.29)&	2.12&	7.8\%&	(1.54,4.60)\\
0.005 (15,0)&	-90.80&	-1.6\%&	(5018.19,6623.66)&	1.08&	4.0\%&	(1.69,4.05)\\
0.006 (15,2.5)&	-110.92&-1.9\%&	(4879.60,6618.11)&	1.42&	5.2\%&	(1.59,4.24)\\
0.012 (15,10)&	-442.60&-7.6\%&	(4130.82,6734.96)&	2.31&	8.5\%&	(1.22,5.05)\\
0.017 (15,15)&	-694.79&-11.9\%&(3762.03,6926.07)&	3.12&	11.5\%&	(1.11,5.60)\\
0.017 (15,-15)&	-257.13&-4.4\%&	(4260.17,7219.80)&	2.08&	7.6\%&	(0.90,5.41)\\\hline    
 \end{tabular}
\end{table}

\begin{figure}[t]
\subfigure[\label{fig:asym:kon}]{
\resizebox{3.1in}{!}{\includegraphics{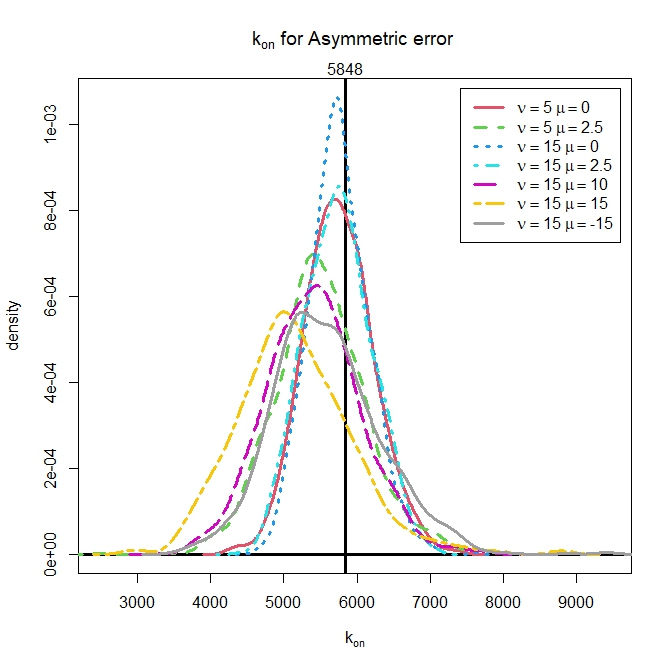}}
}
\subfigure[\label{fig:asym:koff}]{
\resizebox{3.1in}{!}{\includegraphics{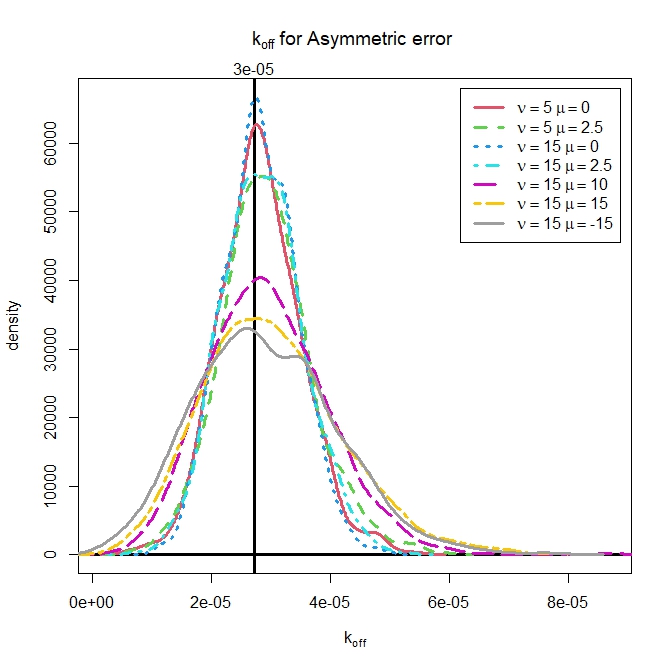}}
}
\caption{Density plots of the estimated $\ka$ and $\km$ from the $1000$ replications of the simulated noisy melting and annealing curves. The standard deviation of the centred $t_{\nu}(\mu)$ noise varies with $\nu$ and $\mu$. 
} 
\label{fig:kAsymErr}
\end{figure}

 The results are presented in Table \ref{tab:asym} and Figure \ref{fig:kAsymErr}.  It follows that $\ka$ is generally negatively biased. The bias depends more on $\mu$, and not much on $\nu$.  It should also be noted that the sign of $\mu$ influences the bias. From Table \ref{tab:asym} it follows that for the same value of $\nu$, the absolute bias for a negative $\mu$ is lower than that of a positive $\mu$. The same seems to be true for $\km$ as well. For both kinetic parameters, the relative biases remain below $10\%$ if the error variance is below $0.01$.       

 Similar results can be obtained for intramolecular hybridisation experiments. The simulation studies shown above prove that the proposed procedure produces accurate results, which should be comparable to any existing procedure for computing the kinetic and thermodynamic parameters of thermal hybridisation experiments.
 
\section{Conclusion}

Statistical tools have been crucial in various chemical analyses for assessing complex experi-mental data, validating results, optimizing experiments, controlling quality, and identifying un-known reaction parameters.  In this article, we describe a statistical procedure to analyse UV thermal curves from nucleic acid hybridization experiments showing hysteresis.

The phenomenon of hysteresis is commonly observed in many nucleic acid hybridisation experi-ments.  It produces broad, irreversible, and non-merging thermal curves, from which the thermo-dynamic parameters of the reaction cannot be directly determined.  Our procedure provides a unified tool to obtain all the kinetic and thermodynamic parameters of the experiment from even hysteric data. Instead of the traditional Savitsky-Golay method, we use a statistical technique called local polynomial regression to optimally smooth the experi-mental data and estimate the values and derivatives of the thermal curves. The kinetic rate con-stants are then computed by solving a pair of linear equations involving the smoothed data and the derivatives. The thermodynamic constants are obtained from these kinetic constants via the Arrhenius plot. The proposed procedure can be implemented using freely available software in the R language.

A user-friendly web-based version of our package is available at https://sanjaychaudhuri.shinyapps.io/anhysnuc/. This package provides reliable inferences from hysteric UV curves, requiring fewer experiments, and thus saving time and effort. It computes the thermodynamic parameters of the process, without requiring additional experiments. This would be otherwise impossible. Additionally, it provides an approximate measure of uncertainty that can be used in quantifying accuracy in nucleic acid (modified or unmodified) research.

\section*{Declaration of Competing Interest}
The authors declare that they have no known competing financial
interests or personal relationships that could have appeared to influence
the work reported in this paper.
\section*{Data and Software Availability} 
A web-based software called {\tt anhysnuc} can be found at \href{https://sanjaychaudhuri.shinyapps.io/anhysnuc/}{https://sanjaychaudhuri.shinyapps.io/anhysnuc/}.  This software implements the whole proposed procedure.  The datasets used in the example are available in the data.zip file uploaded as a supplementary material.
\section*{Acknowledgement}
J. Bhattacharyya acknowledges the partial financial supports from DBT NER-Twinning Project Scheme, and DRDO Govt. of India.  S. Chaudhuri is partially supported by the NSF-DMS grant 2413491 from the National Science Foundation USA.  





\bibliography{statref.bib,chemref.bib}
\end{document}